\documentstyle[epsfig,amsfonts,amsbsy]{elsart}
\let\sectiontmp\section\let\subsectiontmp\subsection \let\appendixtmp\appendix
\def\section{\setcounter{equation}{0}\sectiontmp}
\def\subsection{\subsectiontmp}
\def\theequation{\arabic{section}.\arabic{equation}}
\def\appendix{\def\theequation{\Alph{section}.\arabic{equation}}\appendixtmp}
\input epsf
\def\Tr   {{\rm Tr}}
\def\Re   {{\rm Re}}
\def\Im   {{\rm Im}}
\def\Ks   {\rlap/K}
\def\Ps   {\rlap/P}
\def\Qs   {\rlap/Q}
\def\Rs   {\rlap/R}
\def\Ls   {\rlap/L}
\def\Los  {\rlap/L_1}
\def\Lts  {\rlap/L_2}
\def\Lis  {\rlap/L_i}
\def\Ljs  {\rlap/L_j}
\def\Kps  {\rlap/K_P}
\def\Krs  {\rlap/K_R}
\def\pp   {\mathcal{P}}
\def\(    {\left(}    \def\)   {\right)}
\def\[    {\left[}    \def\]   {\right]}
\makeatletter
\def\ps@copyright{\let\@mkboth\@gobbletwo
  \def\@oddhead{}%
  \let\@evenhead\@oddhead
  \def\@oddfoot{\small\sl
      GSI-Preprint-98-45
	\hfill }
  \let\@evenfoot\@oddfoot
}
\begin{document}
\begin{frontmatter}
\title{Hot $QED$ beyond ladder graphs} 

\author{E. Petitgirard} 

\maketitle

\noindent                        
{\it\small Gesellschaft f\"ur Schwerionenforschung mbH, Planckstr. 1,
64291 Darmstadt, Germany}
\\

\begin{abstract}

At finite temperature a breakdown of the hard thermal loop expansion
arises whenever external momenta are light-like or tend to very soft 
scales. A resummation of ladder graphs is important in these cases where
the effects of infrared or light-cone singularities are enhanced. We show
that in hot $QED$ another class of diagrams is also relevant at leading 
order due to long range magnetic interactions and therefore recent 
studies about ladder expansions need to be corrected. A general
cancellation of the hard modes damping effects still occurs near the 
light-cone or in the infrared region. The validity of an improved
version of the hard thermal loop resummation scheme is discussed.\\ 
\vspace{0.4cm}
\end{abstract}
\end{frontmatter}

\section{Introduction}

The loop expansion for gauge theories at high temperature suffers
from a number of problems due to the extreme nature of the infrared
divergences present. To address these difficulties the hard thermal
loop expansion was devised \cite{r8,klim,wel,brat,wong,tay}.
Although successful in resolving many paradoxes, there still remain 
some fundamental problems with this resummation scheme in certain limits 
outside of its range of validity. One particular problem is that of the 
damping rate of a fast fermion, where a self--consistent calculational scheme 
outside of the hard thermal loop expansion has been used \cite{smilga,piz}. 
Another class of such problems involves processes sensitive to the 
behaviour near the light--cone, where the soft photon production rate 
estimates \cite{baier,brems} already signal a breakdown of this
expansion. In the same context, it was found that a resummation of 
asymptotic thermal masses for hard modes leaves the gauge invariance 
of the effective action intact \cite{flech,kraemmer}. 
\par
But for such processes, a natural question is to know whether or not 
additional resummations beyond those of asymptotic thermal masses are
required. A candidate is the anomalously strong damping of hard modes
from interactions with soft ones. However if damping is to be taken into
account, modifying only the propagators violates the Ward identities and
is therefore not sufficient. Furthermore the damping arising from 
interactions with 
soft modes cannot be incorporated into an effective action which
summarizes the effects of integrating out the hard modes
only. Vertex Corrections are necessary to restore gauge
invariance and leads to a ladder resummation. Perturbation expansions which 
include ladder graphs have been considered in many works. Ladders arise 
for example in the context of the eikonal expansion of gauge theories 
\cite{cornwall,hou}. In the infrared limit of the polarization tensor in 
hot $QED$ \cite{smilga}, but within a simplified model using a constant 
damping, it was claimed that this damping term cancels out from all 
components of $\Pi^{\mu\nu}$. Subsequent studies \cite{kraemmer,car} in scalar
$QED$ have led to the same conclusion for specific limits. More recently
\cite{ckpet,cko} a simple 
and general way to eliminate the damping terms has been put
forward. A justification for such a compensation can be provided by
an argument of gauge-independence. In the estimate of the
damping term, keeping the external line on mass-shell while introducing a
finite infrared cutoff $\mu$ gives a gauge-independent contribution 
\cite{reb}. But with propagators the integration is carried out over the
real axis and usually leaves the momentum off mass-shell. Inserting the
damping term therefore leads to gauge-dependent pieces. For energetic
fermions these pieces are overwhelmed by a
gauge-independent factor $\ln (1/e)$ but problems of
gauge-dependence still remain when going beyond a simple logarithmic 
approximation at leading order. In that respect a general compensation
of the damping terms in the expression of the photon polarization tensor is 
necessary. In short, specific calculations show that the usual hard 
thermal loop term seems to be recovered in the infrared region. 
\par
It is worth recalling that there are simple arguments
based on kinetic equations analysis that suggest the same
conclusion. In the framework of the scalar theory, the effect of the 
infinite set of ladders is shown to be generated by a collision term in
the transport equations \cite{jeon}. Extending this statement to $QED$,
it can be shown that the effect of collisions is suppressed at length
scales $(e^2T)^{-1}$ and starts to become important only at the order
$(e^4T)^{-1}$ \cite{son}.
\par
This study extends the diagrammatic approach of
Refs.~\cite{smilga,kraemmer,ckpet}. The main motivation is to determine
all the relevant diagrams in the light-cone and in 
the infrared region. The main statement is that not only ladders contribute 
at leading order (at least beyond a simple logarithmic approximation), but 
also a full class of graphs with soft photon exchanges. Since the mechanism 
of cancellation is connected to Ward identities, this does not prevent to 
get a final answer in the infrared and in a 'weak' light-cone limit (as it 
will be explained later on). Although the result in these cases is 
expected to be the hard thermal loop, it is useful to study how the
damping terms 
actually disappear. It will be seen that the demonstration 
is not only restricted to specific limits such as the static or the 
zero-momentum limits. It is finally a preparation to investigate the 
light-cone problem where the effects of an asymptotic mass show up. 
Already it demonstrates that the improved resummation scheme proposed in 
Ref.~\cite{flech} seems to 
be justified but might not be complete. Finally it is worth mentioning that 
Refs.~\cite{smilga,kraemmer,ckpet,cko} left technical ambiguities that can 
only be overcome with an improved version of the vertex proposed in this 
paper, in a comparable way to what has been done at zero temperature in 
Ref.~\cite{ball}. The next section presents a study of all the relevant 
graphs, using a precise power counting, the third part is devoted to a 
resummation of these diagrams for the specific cases mentioned above,
and the final demonstration concerns the mechanism of cancellation of
the damping terms.  
\section{Leading diagrams}
\subsection{Ladder graphs}

Several works have been partly devoted to the cancellation of ladder 
graphs in an effective expansion. That has been done either within a 
simplified model, namely with a constant damping, in $QED$ \cite{smilga} or 
scalar $QED$ \cite{kraemmer,car}, or with a momentum-dependent damping 
using algebraic compensations \cite{ckpet,cko}. But it would be
interesting to see to which extent algebraic cancellations survive even 
when taking into account all the leading order diagrams, not only
ladders. That was not the case in \cite{ckpet,cko}. However, before going
beyond a ladder resummation, it is worth recalling, even briefly,
the power counting arguments put forward in these previous studies. Also
it is important to insist on the equivalence between infrared limit and
light-cone limit in this power counting, as the light-cone limit has
just been superficially treated in Ref.~\cite{ckpet}.
\par
Throughout this paper, the retarded/ advanced formalism \cite{aurenche} 
is adopted, more specifically the conventions of Aurenche and
Becherrawy. The structure of Green functions as 'tree-like diagrams' can
easily be seen. Nevertheless all the calculations can be performed
within different real time formalisms, for example the
Schwinger-Keldysh technique \cite{keld}. The simple $ee\gamma$ one-loop 
vertex with $R/A$ prescriptions and the exchange of a soft photon reads
\begin{eqnarray}
\tilde{V}^{\mu}_{RAR} & & (P,Q,-R)=-e^2\int\frac{d^4L}{(2\pi)^4}
 P^t_{\rho\sigma}(L)\( \gamma^{\rho}(\Rs +\Ls )\gamma^{\mu}
(\Ps +\Ls )\gamma^{\sigma}\) \nonumber\\&
&\left\{ \( \frac{1}{2}+n(l_0)\) \( ^*\Delta^t_R(L)
-^*\Delta^t_A(L)\) \Delta_R(P+L)\Delta_A(R+L)\right.\nonumber\\&
&\left.+\( \frac{1}{2}-n_F(p_0+l_0)\) 
\( \Delta_R(P+L)-\Delta_A(P+L)\) \Delta_A(R+L)^*\Delta^t_R(L)\right.
\nonumber\\&
&\left.+\( \frac{1}{2}-n_F(r_0+l_0)\) 
\( \Delta_R(R+L)-\Delta_A(R+L)\) \Delta_R(P+L)^*\Delta^t_A(L)\right\} ,
\end{eqnarray}
where $^*\Delta(L)$ is the effective propagator resummed within the {\it
hard thermal loop} scheme. The particular example of the transverse
component (in covariant gauges) is taken. Each statistical factor is 
associated with a cut internal line and the graph itself appears as a sum of
tree-diagrams. A crucial point (unnoticed in Ref.~\cite{ckpet,cko}) is
that the leading term is given by the Bose-Einstein distribution. 
Splitting the product of fermion propagators $\Delta(P)\Delta(R)$ leads to 
\begin{eqnarray}
\label{premier}
\tilde{V}^{\mu}_{RAR}(P,Q,-R) &=& -e^2\int\frac{d^4L}{(2\pi)^4}
 P^t_{\rho\sigma}(L)\( \gamma^{\rho}(\Rs +\Ls )\gamma^{\mu}
(\Ps +\Ls )\gamma^{\sigma}\) \nonumber\\&
&n(l_0)\rho_T(L)\frac{1}{2Q.(P+L)+Q^2}
\( \Delta_A(R+L)-\Delta_R(P+L)\) .
\end{eqnarray}     
A straightforward estimate can be done when the denominator
$2Q.(P+L)+Q^2$ is of the same order as $2P.Q+Q^2$. Then the leading
vertex has the same magnitude as 
\begin{eqnarray}
\tilde{V}^{\mu}_{RAR}(P,Q,-R) &\sim& -e^2\frac{2P^{\mu}+Q^{\mu}}{2Q.P+Q^2}
\int\frac{d^4L}{(2\pi)^4}
 P^t_{\rho\sigma}(L)\( \gamma^{\rho}\Ps\gamma^{\sigma}\) 
n(l_0)\rho_T(L)\nonumber\\&
&\qquad\qquad\qquad\qquad\qquad\times\( \Delta_A(R+L)-\Delta_R(P+L)\) .
\end{eqnarray}  
Under what circumstances the denominator can be extracted from the
integral remains to be seen. The difference of self-energies written
above has the same order as the damping rate {\it i.e.} $e^2T$
(discarding the well-known problems of logarithmic infrared divergences 
which do no affect the order of magnitude). This is compensated by the 
term $1/(2P.Q+Q^2)$ which brings an extra factor $1/e^2$. The vertex is 
therefore of the same order as its tree-level counterpart. From this power 
counting it was already extensively explained in previous works how the
generation of further ladder graphs leads to the same
order. It can be easily verified from the expressions of 
the multi-loop graphs of Ref.~\cite{aurenche} that the leading 
contributions involve only tree-diagrams with cut photon lines. This 
estimate enables to distinguish between three different limits:
\begin{itemize}
\item The infrared limit where the photon momentum lies in a very soft
      scale, $q_0,\, q\sim O(e^2T)$.
\item A weak light-cone limit for the vertex where the photon is soft 
      $q_0,\, q\sim O(eT)$ but real or almost real $q_0\sim q+O(e^2T)$. 
      If $P$ is one of the electron momenta, the angle of emission is 
      $\hat{p}.\hat{q}\sim \pm 1+O(e)$. 
\item A strong light-cone limit with the same conditions for the photon
      momentum but $\hat{p}.\hat{q}\sim \pm 1+O(e^2)$. This is typically the
      limit of interest for multiple scatterings phenomena. Also if the
      fermion propagators are resummed, the terms involving an
      asymptotic mass contribute \cite{flech}. 
\end{itemize} 
It is important to mention that diagrams with hard photons exchanges do
not contribute, although the phase-space is larger. This is due to
non-trivial compensations between the trace and some denominators. One
of the propagators always gets suppressed and the graphs are not
enhanced as in the previous case. 
\par
It will be seen that ladders with soft photon exchanges cancel against the 
corresponding leading order self-energies without vertex corrections 
(rainbow diagrams). This is due to a mechanism related to Ward
identities. Here the status concerning the spectral density has not been 
precised. Unlike the damping rate problem the external momentum of the
self-energy is a variable. Kinematics does not forbid the exchanges of
time-like photons, even though it is the Landau damping part of the
spectral density which should give the dominant contribution, the
fermion is still close to its mass-shell. Now with the Landau damping
part the photon momentum can reach the infrared scale $e^2T$ due to the
absence of magnetic screening and this may change the estimate of
multi-loop diagrams. If it turns out that self-energies with vertex 
corrections {\it at leading order} contribute too, then graphs other 
than ladders might be relevant for consistency and therefore it is 
necessary to go beyond what has been done in 
\cite{smilga,kraemmer,ckpet,cko}. 
\subsection{Two-loop self-energy}

The two-loop retarded self-energy for an on-shell electron is
considered. Since the exchange of transverse photons is characterized by
the absence of (static) magnetic screening, a naive power counting
suggests that the infrared scale of $O(e^2T)$ could dramatically
change the order of higher loop diagrams. This is due to power-like
infrared divergences and, as a first example, the simple vertex correction
of the self-energy might contribute at the same leading order as its
bare counterpart. Simple estimates made in \cite{smilga} within a
simplified model led to the conclusion that such vertex corrections are
subleading, while it has been pointed out in \cite{blaiz} that these
corrections conspire to give a leading order evaluation of the damping
rate. There is a need to clarify the situation, since the presence of
leading vertex corrections in the self-energies would imply that ladders
are not the only relevant diagrams in the infrared and light-cone limits
of the polarization tensor. 
\par
A straightforward estimate of this two-loop graph can be obtained using 
the simplified transverse spectral density introduced in 
Ref.~\cite{blaiz}
\begin{equation}
\label{approspectral}
\frac{\rho_T(Q)}{q_0} = \frac{2\pi}{q^2}\delta(q_0) .
\end{equation}
The complete calculation, starting with the exact transverse density, 
will be presented in the appendix. The retarded self-energy in the $R/A$
formalism \cite{aurenche} can be written as
\begin{eqnarray}
-i\Sigma_{RR}(P)& =& -e^2\int\frac{d^4Q}{(2\pi)^4}n(q_0)\Delta_R(P+Q)
   P^t_{\mu\nu}(Q)\[
   \tilde{V}^{\mu}_{ARA}(-Q,P+Q,-P)^*\Delta_R(Q)\right.\nonumber\\&
  &\left. -\tilde{V}^{\mu}_{RRA}(-Q,P+Q,-P)^*\Delta_A(Q)\] (\Ps +\Qs )
 \gamma^{\nu},
\end{eqnarray} 
with the expression 
\begin{eqnarray}
\tilde{V}^{\mu}_{ARA}(-Q,P+Q, & & -P)=
-e^2\int\frac{d^4K}{(2\pi)^4}n(k_0)\rho_T(K)P^t_{\rho\sigma}(K)
\Delta_R(P+Q+K)\nonumber\\&
&\Delta_R(P+K)\gamma^{\rho}(\Ps +\Ks )\gamma^{\mu}
(\Ps+\Qs +\Ks )\gamma^{\sigma}.
\end{eqnarray}
In this specific case the $R/A$ indice of the photon does not modify the
expression of the vertex
\begin{equation}
\tilde{V}^{\mu}_{ARA}(-Q,P+Q,-P)= \tilde{V}^{\mu}_{RRA}(-Q,P+Q,-P).
\end{equation}
A direct evaluation of the order of magnitude can be obtained if the
self-energy itself is replaced by a simpler quantity without modifying
the order. Such a quantity may be the imaginary part of the trace 
$\Tr \( \Ps \Sigma(P)\) $. In order to simply evaluate the order it
always makes sense to take $P^2=0$. Next, the usual simplifications 
can be made, {\it i.e.} keeping only hard terms in the trace and 
using the approximate Bose-Einstein factors $T/q_0$ and $T/k_0$ for 
soft photons. Finally the damping of a particle with positive energy is
chosen. Retaining only the positive energy parts of the propagators gives
\begin{eqnarray}
\label{zetaplus}
\frac{1}{4p_0}\Tr \( \Ps \Sigma_{RR}(P)\)   
 & =& (e^2T)^2\int\frac{d^3q}{(2\pi)^3}\int\frac{dq_0}{(2\pi)^3q_0}\rho_T(Q)
      \int\frac{d^3k}{(2\pi)^3}\int\frac{dk_0}{(2\pi)^3k_0}\rho_T(K)
      \nonumber\\&
 &\Im \frac{1}{q_0-\hat{p}\cdot\vec{q}+i\epsilon}\:
      \frac{1}{k_0-\hat{p}\cdot\vec{k}+i\epsilon}\:
      \frac{1}{q_0+k_0-\hat{p}\cdot(\vec{q}+\vec{k})+i\epsilon}\nonumber\\&
&\( 1-(\hat{p}\cdot\hat{q})^2\) \( 1-(\hat{p}\cdot\hat{k})^2\) .
\end{eqnarray}
where the spectral density depicting the exchange of static magnetic 
photons is given by the simplified form of Eq.~[\ref{approspectral}]. 
Upper and lower cut-offs need to be introduced by hand. These can be
respectively the plasmon frequency $\omega_p$ and the parameter $\mu$, 
the former being of order $eT$ and the latter of order $e^2T$. Taking 
into account the non trivial phase space due to kinematical constraints,
it turns out that the expression of Eq.~[\ref{zetaplus}] can be
decomposed into three parts
\begin{eqnarray}
A_1& =& -2i\pi(e^2T)^2\int_{\mu}^{\omega_p}\frac{dq}{(2\pi)^2q^2}
\int_{-1}^{1}dx\( \frac{\pp}{x^2}-1\) \int_{\mu}
^{\omega_p}\frac{dk}{(2\pi)^2k}\nonumber\\&
     =& \frac{4i}{(2\pi)^3}(e^2T)^2\frac{1}{\mu}\ln\( \frac{\omega_p}{\mu}\) 
\end{eqnarray}
\begin{eqnarray}
A_2& =& i\pi(e^2T)^2\int_{\mu}^{\omega_p}\frac{dq}{(2\pi)^2q^2}
\int_{-1}^{1}dx\( \frac{\pp}{x^2}-1\) \int_q^{\omega_p}\frac{dk}{(2\pi)^2k} 
   \int_{-1}^1dy\delta\( y+\frac{q}{k}x\) 
  \( 1-y^2\) \nonumber\\&
     =& \frac{-2i}{(2\pi)^3}(e^2T)^2\frac{1}{\mu}\( \ln\(
\frac{\omega_p}{\mu}\) -\frac{5}{6}\) 
\end{eqnarray}
\begin{eqnarray}
A_3& =& i\pi(e^2T)^2\int_{\mu}^{\omega_p}\frac{dq}{(2\pi)^2q^2}
   \int_{\mu}^{q}\frac{dk}{(2\pi)^2k}
   \int_{-\frac{k}{q}}^{\frac{k}{q}}dx\( 
    \frac{\pp}{x^2}-1\) 
  \int_{-1}^1dy\delta\(
   y+\frac{q}{k}x\) \( 1-y^2\) \nonumber\\&
     =& \frac{-2i}{(2\pi)^3}(e^2T)^2\frac{1}{\mu}\( \ln\(
   \frac{\omega_p}{\mu}\) -\frac{5}{6}\)  
\end{eqnarray}
A cancellation of the leading singularity $(1/\mu)\ln(\omega_p/\mu)$ 
occurs between these three parts. Still a term $1/\mu$ does not get 
eliminated
\begin{equation}
\frac{1}{4p_0}\Tr \( \Ps \Sigma_{RR}(P)\) = A_1+A_2+A_3
= (e^2T)^2\frac{10i}{3(2\pi)^3}\frac{1}{\mu}.
\end{equation}
This means that, when dealing with complete fermion propagators instead
of bare ones, dressed by full self-energies $\Sigma(P+L_i)$, the diagram 
is sensitive to the scale $e^2T$ with $\Sigma(P+L_i)$ providing possible
cut-offs. In conclusion, the two-loop graph is at the same
leading order $e^2T$ as the one-loop self-energy. Actually multi-loop 
vertex corrections must be taken into account whenever transverse
photons are considered. 
\subsection{Non-planar diagram}

The importance of a ladder resummation in the infrared limit is a
well-known fact. Since the cancellation of the damping terms is related
to Ward identities, ladders by themselves correspond to the elimination of
self-energies without vertex corrections. However it has been noticed in
the previous section that vertex corrections contribute at the 
same leading order. Therefore diagrams in the $ee\gamma$ vertex
strictly related to these corrections via Ward identities must also be
taken into account. In particular, non-planar graphs are shown to be
relevant contrary to what has been claimed in  Ref.~\cite{ckpet} and 
repeated in ref.~\cite{cko}. 
\par
In \cite{ckpet,cko} the example of a higher loop crossed
graph is mentioned. The case of a scalar $\lambda\phi^3$ theory at zero
temperature is considered but with a generalization for other theories 
at finite temperature in sight. Using the same notations as in
\cite{ckpet,cko} the two-loop non-planar diagram is written as 
\begin{eqnarray}
  -i\Pi(K)& =& (-i\lambda)^6\int\, dR_1\, dR_2\, dP\,
  \Delta(P+R_1+R_2)\Delta(P+R_1+R_2+K)\Delta(R_1)\nonumber\\
  & &\Delta(P+R_2)\Delta(P+R_1+K)\Delta(R_2)
  \Delta(P)\Delta(P+K),
\end{eqnarray}
where $\Delta(K)=1/(K^2+i\epsilon)$. It is claimed that this kind of graphs
does not contribute in the same way as the ladder graphs. While
the product of propagators $\Delta(P+R_1+R_2)\Delta(P+R_1+R_2+Q)$ would
be split in the infrared limit $2Q\cdot R_1+R_2\ll Q^2+2Q\cdot P$ as
\begin{eqnarray}
  \Delta(P+R_1+R_2) & & \Delta(P+R_1+R_2+Q)\nonumber\\&
& =i\frac{\Delta(P+R_1+R_2)-\Delta(P+R_1+R_2+Q)}{Q^2+2Q\cdot P},
\end{eqnarray}
splitting the product $\Delta(P+R_2)\Delta(P+R_1+Q)$ in the same limit 
$2Q\cdot R_1 \ll Q^2 + 2Q\cdot P$ then would produce
\begin{equation}
  \Delta(P+R_2)\Delta(P+R_1+Q) \approx
  i\frac{\Delta(P+R_2)-\Delta(P+R_1+Q)}
  {Q^2+2Q\cdot P + (P+R_1)^2 - (P+R_2)^2}.
\end{equation}
By itself it would not lead to a cancellation of a factor of $\lambda^2$ 
in the numerator due to the presence of the $(P+R_1)^2 - (P+R_2)^2$ term.
The argument put forward is that even by furthermore restricting the 
phase space so that $P\cdot R_i$ and $R_i^2$ ($i=1,2$) is sufficiently 
small, this introduces extra factors of $\lambda$ in the numerator
coming from the momentum integral over $P$. The conclusion drawn is that
in the infrared limit such crossed graphs are suppressed relative to the
ladder graphs. 
\par 
However the point is that, when considering transverse spectral
densities in gauge theories there is no restriction of the phase-space 
(extra factors $\lambda$ in the numerator) because of the absence of 
magnetic screening at the order $\lambda T$. This is responsible for 
the well-known infrared sensitivity of the hard fermion damping rate. 
Therefore terms like $P\cdot R_i$ and $R_i^2$ ($i=1,2$) can be
sufficiently small without rendering the 
corresponding regions of integration negligible. More precisely, a
power counting can be provided for this crossed diagram. Only
bare fermion propagators are considered at this stage. Soft photons 
propagators are resummed within the $HTL$ scheme as they should, 
and the approximate spectral density of Eq.~[\ref{approspectral}] 
may be used. The basic expression of the crossed graph in the $R/A$
formalism is already complicated. But it can be found that the leading
pieces correspond to the tree-like terms with cut internal photon
lines. The dominant contribution of the crossed diagram with the $RAR$ 
prescriptions as a particular example, is 
\begin{eqnarray}
\tilde{V}_{RAR}^{\mu}(P,Q,-R) &=&
e^4\int[dL_1]^t_{\alpha\beta}[dL_2]^t_{\sigma
\rho}\gamma^{\alpha}(\Rs +\Los )\Delta_A(R+L_1)\gamma^{\sigma}(\Rs +\Los
+\Lts )\nonumber\\&
&\Delta_A(R+L_1+L_2)\gamma^{\mu}(\Ps +\Los +\Lts )\Delta_R(P+L_1+L_2)
\nonumber\\&
&\gamma^{\beta}(\Ps +\Lts )\Delta_R(P+L_2)\gamma^{\rho},
\end{eqnarray}
where the shorthand $[dL_i]$ stands for
\begin{equation}
\label{intdensit}
[dL_i]^t_{\alpha\beta}=\frac{d^4L_i}{(2\pi)^4}n(l_{i_0})\rho_T(L_i)
P^t_{\alpha\beta}(L_i).
\end{equation} 
The usual decomposition of the fermion propagator reads 
\begin{equation}
\Delta_R(P)=\frac{i}{2\Omega_P}\( \frac{1}{p_0-\Omega_P+i\epsilon}-\frac{1}
{p_0+\Omega_P+i\epsilon}\) .
\end{equation}
As it is only a matter of determining the order of magnitude of a graph,
various simplifications can be made. First the first term of the
decomposition above corresponds to an electron propagating forward in 
time, and the second term to an antiparticle also propagating forward in
time. All the quantities involve soft photons and are related to the
picture of an energetic electron (or positron) close to its mass-shell 
undergoing multiple interactions with soft photons. The anihilation
process between a particle and an antiparticle is subleading compared to
simple scattering processes of the same particle. In the expression of
the vertex above only the positive energy parts of the propagators may
be retained, in order to find out just the order of magnitude. This
stricly follows the lines of the previous estimate concerning the
two-loop self-energy diagram. Splitting the product 
$\Delta(P+L_1+L_2)\Delta(R+L_1+L_2)$ and keeping only the
positive energy components give the term
\begin{eqnarray}
\label{crossbare}
\tilde{V}^{+^{\mu}}_{RAR}(P,Q, & & -R) =
e^4\int[dL_1]^t_{\alpha\beta}[dL_2]^t_{\sigma
\rho}\gamma^{\alpha}(\Rs +\Los )\Delta^+_A(R+L_1)\gamma^{\sigma}(\Rs +\Los
+\Lts )\nonumber\\&
&\frac{1}{q_0+\Omega_{P+L_1+L_2}-\Omega_{R+L_1+L_2}}\( \Delta^+_R(P+L_1+L_2)
-\Delta^+_A(R+L_1+L_2)\) \nonumber\\&
&\gamma^{\mu}(\Ps +\Los +\Lts )\gamma^{\beta}(\Ps +\Lts )
\Delta^+_R(P+L_2)\gamma^{\rho},
\end{eqnarray} 
where the $i\epsilon$ prescription is absorbed into a re-definition of
$q_0$. It will become clear in the next sections that the same procedure is
employed when dealing with resummed propagators. The common denominator 
$1/(q_0+\Omega_{P+L_1+L_2}-\Omega_{R+L_1+L_2})$ has the same order as 
$1/(q_0-\Omega_P+\Omega_R)$. Under what circumstances this replacement 
can actually be done will be seen in details later on. The last step of
the power counting consists in taking particular terms easily calculable
under specific conditions from expression [\ref{crossbare}]. Such a 
procedure should not change the order of magnitude. This is the case
when taking the imaginary part $(1/4p_0)\Im \Tr (\Ps \tilde{V})$
evaluated with both $P$ and $R$ on-shell (internal momenta are close to
their mass-shell). It turns out that
\begin{eqnarray}
\tilde{V}^{+^{\mu}}_{RAR}(P,Q,-R) \sim  \frac{\hat{P}^{\mu}}
{q_0+\Omega_P-\Omega_R}\(
\frac{1}{4p_0}\Im\Tr (\Ps\Sigma_1(P))
-\frac{1}{4p_0}\Im\Tr (\Ps\Sigma_2(P))\) ,\nonumber\\
\end{eqnarray}
where 
\begin{eqnarray}
\frac{1}{4p_0}\Im\Tr (\Ps\Sigma_1(P)) &=& -(e^2T)^2\int\frac{d^3l_1}
{(2\pi)^3}\int\frac{d^3l_2}{(2\pi)^3}\( 1-(\hat{p}\cdot\hat{l_1})^2\) 
\( 1-(\hat{p}\cdot\hat{l_2})^2\) \nonumber\\&
&\Im \frac{1}{\hat{p}\cdot\vec{l_1}+i\epsilon}\:
      \frac{1}{\hat{p}\cdot\vec{l_2}-i\epsilon}\:
      \frac{1}{\hat{p}\cdot(\vec{l_1}+\vec{l_2})-i\epsilon}.
\end{eqnarray}  
Thereby $(1/4p_0)\Im\Tr (\Ps\Sigma_2(P))$ has the same expression 
except a $+i\epsilon$ prescription in the last denominator, and the sign
$\sim$ means 'of the same order as'. In the same way as with the two-loop 
self-energy, the infrared cut-off $\mu \sim O(e^2T)$ must be introduced.
The term $1/\mu$ does not get canceled
\begin{equation}
\tilde{V}^{+^{\mu}}_{RAR}(P,Q,-R)\sim\frac{\hat{P}^{\mu}}
{q_0+\Omega_P-\Omega_R}
(e^2T)^2\frac{2}{(2\pi)^3}\frac{1}{\mu}\ln\( \frac{\omega_p}{\mu}\) ,
\end{equation}
and shows the sensitivity to the infrared regime (quantities 
$\sim O(e^2T)$). With resummed fermion propagators the crossed diagram
will be of the same order as the ladders. The inclusion of higher order
crossed graphs is necessary. This is consistent with the fact that
vertex corrections in the self-energies are relevant at leading order. 
\subsection{Resummed fermion propagator}

It is important at this stage to review all the relevant diagrams
involved in the self-energy. All these graphs have to be resummed to
form the propagator of a hard fermion close to its mass-shell. First, as
advocated in Ref.~\cite{flech}, the {\it hard thermal loop} self-energy
must be considered. The well-known two-structure functions of the $HTL$
term read
\begin{equation}
\Sigma_{HTL}(P)=ap_0\gamma_0+b\vec{p}.\vec{\gamma},
\end{equation}
where
\begin{equation}
a=-\frac{e^2T^2}{2pp_0}\ln\( \frac{p_0+p}{p_0-p}\) ,\qquad
b=\frac{e^2T^2}{p^2}\[ 1-\frac{p_0}{2p}\ln\( \frac{p_0+p}{p_0-p}\) \] .
\end{equation}
Also the self-energy $\Sigma$ involving soft longitudinal and 
transverse photon exchanges must be resummed. The dressed (for example
retarded) propagator is given by
\begin{eqnarray}
S_R(P) & =&\frac{i}{\Ps -\Sigma_{HTL}(P)-\Sigma_R(P)}\nonumber\\
&=&  \frac{i\Ps +O(e^2)}{P^2-2ap_0^2-2bp^2-\Ps \Sigma_R(P)
-\Sigma_R(P)\Ps }.
\end{eqnarray}
The combination of the $HTL$ structure functions gives the asymptotic
mass \cite{flech}
\begin{equation}
m_{\infty}^2=2(p_0^2a+p^2b)=\frac{1}{4}e^2T^2.
\end{equation}
It is also necessary to introduce (with implicitly the same definition
for the advanced counterpart)
\begin{equation}
\sigma_R(P)=-\frac{1}{4p_0}\Tr \( \Ps \Sigma_R(P)\) ,
\end{equation}
such that the imaginary part of this quantity just corresponds to the
damping rate. Although the real part might be subleading, it is better
to keep this real term in the following. It will get eliminated along
with the damping in a general mechanism of cancellation. The fermion 
propagator finally reads
\begin{equation}
\label{completeprop}
S_R(P)=\frac{i\Ps +O(e^2)}{P^2-m_{\infty}^2+2p_0\sigma_R(P)},
\end{equation}
where it is worth noting that without $\sigma$ (and damping resummation)
the propagator introduced in Ref.~\cite{flech} is obviously recovered.
The 'one loop' contribution to $\Sigma$ is
\begin{equation}
\Sigma^1_R(P)=\sum_{i=t,l}(-ie^2)\int[dL]^i_{\alpha\beta}\gamma^{\alpha}
S_R(P+L)\gamma^{\beta} ,
\end{equation}
where in fact the resummed propagator is not necessary in the
longitudinal part. There is no sensitivity to the infrared scale due to Debye
screening. The 'two-loop' self-energy involves just the simple vertex
correction with transverse exchanges
\begin{equation}
\Sigma^2_R(P)=-ie^2\int[dL]^t_{\alpha\beta}\tilde{V}^{\alpha}_{ARR}
(-P,L,P+L)S_R(P+L)\gamma^{\beta} .
\end{equation}
It is therefore the Landau damping part of the spectral density which
gives the dominant piece here. The time-like part leads to
subleading terms. That will be the same for the multi-loop vertices. 
However for the ladders and the 'one-loop' self-energy
kinematics does not forbid completely processes with (time-like)
quasi-particles. In order to simplify the notations the entire
expression of the transverse spectral density may be kept for all the 
diagrams, thus including negligible terms in most of the cases. This
will be implicit for higher-order vertices. The following 'N-loop' 
self-energies correspond to all the possible vertex
corrections (with the same resummed fermion propagator). There are four
'three-loop' graphs, and by the multiple insertions of a photon leg
along the fermion line, each of these graphs is related to a subset of
'N-loop' $ee\gamma$ vertices (always with the same fermion propagator) 
via Ward identities. The complete self-energy introduced above may be 
written as
\begin{equation}
\Sigma_R(P)=\Sigma_R^1(P)+\Sigma_R^2(P)+\sum_{K=1}^4\Sigma_{K_R}^3(P)+...
\end{equation}
\section{Ladders and diagram resummation}
\subsection{Preliminaries}

In this section a method is described for including all the diagrams
discussed previously in an effective expansion. Unlike what has been
done before \cite{smilga,ckpet,cko}, this does not only involve ladders 
but all the loop graphs of the four-fermion amplitude which cannot be
disconnected by cutting two fermion lines. These contain resummed
fermion propagators as well as soft photon propagators (longitudinal 
and transverse for the ladders, transverse for all the other cases). A
first non-trivial example was the subset of 'two-loop' graphs, namely
the crossed diagram with the two symetric vertex corrections. The next
subset of 'three-loop' diagrams for example is determined from the four
'three-loop' self-energies via Ward identities ({\it i.e.} via the 
insertions of an external photon leg). This can be summarized in the 
Bethe-Salpeter equation with an infinite kernel
\begin{eqnarray}
\label{betsal}
\tilde{V}^{\mu}(P,Q,-R) &=& \gamma^{\mu}+\int\frac{d^4P'}{(2\pi)^4}
 K(P,-P',R',-R)\Krs \Rs'\Delta(R')\( \gamma^{\mu}+\tilde{V}^{\mu}(P',Q,-R')\)
\nonumber\\&
&\Ps'\Delta(P')\Kps ,
\end{eqnarray}
where
\begin{eqnarray}
K(P,-P',R',-R)\Kps \cdot \Krs &=& K^1(P,-P',R',-R)\Kps^1 \cdot \Krs^1
\nonumber\\&
&+K^2(P,-P',R',-R)\Kps^2 \cdot \Krs^2 +...\nonumber\\&
&R=P+Q;\qquad R'=P'+Q.\nonumber 
\end{eqnarray}
In this equation, the kernel $K$ represents the infinite series of 
four-electron amplitudes. $K^1$ corresponds to all the simple ladders,  
$K^2$ to the crossed and vertex correction graph resummation, etc...
\par
Previous works on the ladders \cite{smilga,kraemmer,ckpet} have led to
the following formula for the vertex
\begin{equation}
  \label{qedrelation}
  \tilde{V}_{\mu}(P,Q,-R) \approx \frac{2P_\mu + Q_\mu}{Q^2+2Q\cdot P}
  \[ \Sigma(P) -\Sigma(R)\] ,
\end{equation}
where the $R/A$ prescriptions have been omitted, and the self-energies
do no include vertex corrections. Although this expression corresponds
to a ladder summation, it is not well-defined due to the absence of
an $i\epsilon$ prescription or a cut-off in the denominator. In particular
sticking this form  into the polarization tensor does not solve the
problem and leaves the same ambiguity \cite{kraemmer,ckpet}. In order to
guess the correct formula for the vertex a good starting point is to
consider the one loop graph with bare propagators in a $\lambda\phi^3$
theory
\begin{equation}
\tilde{V}(P,Q,-R) = -\lambda^2\int[dL]\Delta(R+L)\Delta(P+L).
\end{equation}
The propagator can be split into a positive energy part and a negative
one. Thereby the product $\Delta(P+L)\Delta(R+L)$ leads to subleading
denominators for mixed (positive/ negative energies) terms. The vertex
now reads
\begin{eqnarray}
\tilde{V} & & (P,Q,-R)\approx \lambda^2\int[dL]
\[  \frac{1}{q_0+\Omega_{P+L}-\Omega_{R+L}}\(
\frac{1}{p_0+l_0-\Omega_{P+L}}-\frac{1}{r_0+l_0-\Omega_{R+L}}\) 
\right.\nonumber\\&
&\left.-\frac{1}{q_0+\Omega_{R+L}-\Omega_{P+L}}\(
\frac{1}{p_0+l_0+\Omega_{P+L}}-\frac{1}{r_0+l_0+\Omega_{R+L}}\) \] 
\frac{1}{4p^2}
\end{eqnarray}
Anticipating the fact that under specific circumstances, especially the
infrared limit, denominators like $q_0+\Omega_{P+L}-\Omega_{R+L}$ can be
approximated by $q_0+\Omega_P-\Omega_R$ and therefore can be extracted 
from the integral, finally yields
\begin{eqnarray}
\label{simpleform}
\tilde{V}(P,Q,-R) &=& \frac{1}{q_0+\Omega_P-\Omega_R}\(
\Sigma^+(P)-\Sigma^+(R)\) \nonumber\\&
& \qquad\qquad\qquad\qquad+\frac{1}{q_0+\Omega_R-\Omega_P}
  \( \Sigma^-(P)-\Sigma^-(R)\) ,
\end{eqnarray}
where
\begin{equation}
\Sigma^{\pm}(P)= \pm
\lambda^2\int[dL]\frac{1}{4p^2}\:\frac{1}{p_0+l_0
\mp\Omega_{P+L}}.
\end{equation}
The common denominators $q_0\pm\Omega_P\mp\Omega_R$ no longer include 
$p_0$ and if an integration is performed over this variable only the
part of the simplified propagators is concerned. Now it is not difficult
to guess  a well-defined form for the full $ee\gamma$
vertex. Defining the vectors 
\begin{equation}
v^{\mu}=\( 1,\( 1-\frac{m^2_{\infty}}{2p^2}\) \hat{p}^i\) ,\qquad
\bar{v}^{\mu}=\( 1,-\( 1-\frac{m^2_{\infty}}{2p^2}\) \hat{p}^i\) . 
\end{equation}
it can be written as 
\begin{eqnarray}
\label{completeform}
\tilde{V}_{RAR}^{\mu}(P,Q,-R) &=& \gamma^{\mu}+
\frac{v^{\mu}}{q_0+\Omega_P-\Omega_R}\(
\Sigma^+_R(P)-\Sigma^+_A(R)\) \nonumber\\&
&\qquad\qquad\qquad\qquad
+\frac{\bar{v}^{\mu}}{q_0+\Omega_R-\Omega_P}\( 
\Sigma^-_R(P)-\Sigma^-_A(R)\) .
\end{eqnarray}
A justification {\it a posteriori} will be given in the next
sections. It will be seen that the positive and negative energy
components of the propagators are relevant
\begin{equation}
\Delta^+_{R/A}(P)=\frac{i}{2p}\:
\frac{1}{p_0-\Omega_P+\sigma_{R/A}^+(P)},
\qquad\Delta^-_{R/A}(P)=-\frac{i}{2p}\:
\frac{1}{p_0+\Omega_P+\sigma_{R/A}^-(P)}
\end{equation}
and replace the full expression of the fermion propagator in $\Sigma^+$
and $\Sigma^-$ respectively at any order. These 'complete' self-energies
imply also resummed photon propagators as well as vertex 
corrections. Also the external photon energy is understood to be complex
$q_0=\Re\, q_0+i\epsilon$ since the retarded photon prescription must be
kept ($q_0=\Re\, q_0-i\epsilon$ for an 'advanced' photon). The four-momenta
$v^{\mu}$ and $\bar{v}^{\mu}$ are slighty 
different versions of the unit vectors $(1,\hat{p}^i)$ and
$(1,-\hat{p}^i)$. The subleading terms involving the asymptotic mass
have been introduced in order to fully satisfy Ward identities for the
expression above. It must be understood that this addition is unimportant
as long as calculations are performed at leading order.  
\subsection{Ladder resummation}

\begin{figure}
\begin{center}
\leavevmode
\epsfxsize=4 in
\epsfbox{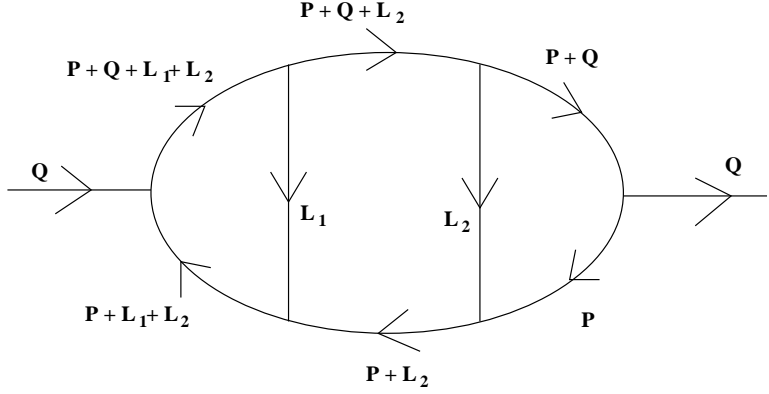}
\end{center}
\caption{A three--loop self--energy ladder graph. The photon
         ($L_1$ and $L_2$) and electron propagators are understood to be
          resummed.}
\label{ladderfig}\end{figure}

In previous works \cite{smilga,kraemmer,ckpet} a resummation of 
ladders has been considered, discarding the possible inclusion of more
complicated diagrams (see Fig.~\ref{ladderfig}). Defering the discussion
about these diagrams to the following sections, the first step consists 
in including the expression  of Eq.~[\ref{completeform}] in the first 
term of the Bethe-Salpeter equation corresponding to a ladder
summation. The complete form of the vertex [\ref{completeform}] should 
lift any ambiguity concerning the absence of the $i\epsilon$
prescriptions in denominators such as $1/P.Q$ and enables to get general
expressions in the infrared, not just specific limits like
$\Pi^{00}(q_0,0)$ and $\Pi^{ii}(0,q\rightarrow 0)$.  
\par
Sticking the expression [\ref{completeform}] into the first 
term of Eq.~[\ref{betsal}] yields
\begin{eqnarray}
\tilde{V}^{1^{\mu}}_{RAR}(P,Q,-R) &=& \sum_{i=t,l}(-e^2)\int[dL]^i_{\rho\sigma}
\Delta_R(P+L)\Delta_A(R+L)\gamma^{\rho}(\Rs +\Ls )
\nonumber\\&
&\[ \gamma^{\mu} +\tilde{V}^{\mu}_{RAR}(P+L,Q,-R-L)\] (\Ps +\Ls )
\gamma^{\sigma},
\end{eqnarray}
where the $RAR$ indices have been chosen as a particular example, the
generalization to any kind of $R/A$ vertex being straightforward. The
indice $1$ is a reminder that the vertex is a 'one-loop' graph with
resummed fermion propagators and dressed transverse and longitudinal 
photons. First it is useful to define
\begin{equation}
\label{sigmapm}
\sigma_{R/A}^{\pm}(P)=-\frac{1}{4p_0}\Tr \( \Ps \Sigma_{R/A}^{\pm}(P)\) ,
\qquad\qquad\Sigma(P)=\Sigma^+(P)+\Sigma^-(P).
\end{equation}
Contracting the expression of the vertex with the spinors leads to 
\begin{eqnarray}
(\Rs +\Ls )\gamma^{\mu}(\Ps +\Ls ) &= & 2(P^{\mu}+L^{\mu})
(\Rs +\Ls)-(\Rs +\Ls )(\Ps+\Ls )\gamma^{\mu}\nonumber\\
& \simeq & 2P^{\mu}\Ps ,\nonumber\\
(\Rs +\Ls )v^{\mu}\Sigma^+(P+L)(\Ps +\Ls ) &= & \frac{1}{2}v^{\mu}\Tr 
\[ (\Ps +\Ls )\Sigma^+(P+L)\] (\Rs +\Ls )\nonumber\\&
&\qquad -v^{\mu}(\Rs +\Ls )(\Ps+\Ls )\Sigma^+(P+L)\nonumber\\
& \simeq & -2P^{\mu}\sigma^+(P)\Ps ,
\end{eqnarray}
where soft terms may be discarded. Furthermore the phase-space is
limited to the regions where the electrons are close to their mass-shell
($P^2\sim O(e)$ at most) which justifies some of the approximations made
above. This can be seen on complex analysis grounds. Replacing the
damping terms of the propagators by simple $i\epsilon$ prescriptions, 
the discontinuities are simple poles. Picking up these 
poles involve the constraints $p_0+l_{i_0}\simeq\Omega_{P+L_i}$ or
$p_0+l_{i_0}\simeq-\Omega_{P+L_i}$ for a specific fermion momentum. This
remains true for the other momenta whose differences are given by the soft
terms $L_i$. These approximate conditions should 
not be modified when coming back to the resummed propagators. This 
argument will be repeated throughout this paper. The expression written 
above now reads (with still a complex energy $q_0=\Re\, q_0+i\epsilon$) 
\begin{eqnarray}
\label{intermvert}
\tilde{V}^{1^{\mu}}_{RAR} & & (P,Q,-R)=\sum_{i=t,l}(-e^2)
\int[dL]^i_{\rho\sigma}
\Delta_R(P+L)\Delta_A(R+L)2P^{\mu}\gamma^{\rho}\Ps \gamma^{\sigma}
\nonumber\\&
&\[ 1-\frac{\sigma_R^+(P+L)-\sigma_A^+(R+L)}{q_0+\Omega_{P+L}-\Omega_{R+L}}
-\frac{\sigma_R^-(P+L)-\sigma_A^-(R+L)}{q_0+\Omega_{R+L}-
\Omega_{P+L}}\] .
\end{eqnarray}
The fermion propagator may be decomposed into a positive energy  
and a negative energy part. As explained above the contribution 
of the resummed vertex in the infrared and on the light-cone becomes 
relevant only when the momenta $P+L_i$ approach their mass-shell, either 
with a positive energy or a negative one, namely
$p_0\simeq\pm\Omega_P$. That enables to split the entire expression of 
the vertex into a positive and a negative part. Under these 
conditions, both the self-energy $\sigma^+$ inside the positive
component of the propagator and $\sigma^-$ inside the negative one
become negligible. It turns out that
\begin{eqnarray}
\Delta_{R/A}(P) &=& \frac{i}{P^2-m_{\infty}^2+2p_0\sigma_{R/A}(P)}
\nonumber\\
&\simeq & \frac{i}{2p}\( \frac{1}{p_0-\Omega_P+\sigma_{R/A}^+(P)}
-\frac{1}{p_0+\Omega_P+\sigma_{R/A}^-(P)}\) \nonumber\\
&=& \Delta^+_{R/A}(P)+\Delta^-_{R/A}(P).
\end{eqnarray}
Considering  the product of propagators $\Delta_R(P)\Delta_A(R)$ the 
previous decomposition leads to
\begin{eqnarray}
\Delta_R(P)\Delta_A(R) &\simeq & \frac{i}{2p}\( \frac{1}{q_0+\Omega_P-
\Omega_R-\sigma_R^+(P)+\sigma_A^+(R)}\[ \Delta^+_R(P)-\Delta^+_A(R)
\] \right.\nonumber\\&
&\left.- \frac{1}{q_0+\Omega_R-\Omega_P+\sigma_A^-(R)-\sigma_R^-(P)}\[
\Delta^-_R(P)-\Delta^-_A(R)
\] \) , 
\end{eqnarray}
where crossed (positive times negative) terms exhibit subleading common 
denominators. For the same reason as above when the momenta $P+L_i$ are
almost on-shell, with a positive energy (resp. negative) the
self-energies $\sigma^-$ (resp. $\sigma^+$) are subleading. This gives
for instance 
\begin{eqnarray}
& & \[ \Delta^+_R(P)- \Delta^+_A(R)\] \[ 1-\frac{\sigma_R^+(P)-
\sigma_A^+(R)}{q_0+\Omega_P-\Omega_R}
-\frac{\sigma_R^-(P)-\sigma_A^-(R)}{q_0+\Omega_R-\Omega_P}\]
\nonumber\\&
&\qquad\qquad\simeq\[ \Delta^+_R(P)-\Delta^+_A(R)\] \[ 1-\frac{\sigma_R^+(P)-
\sigma_A^+(R)}{q_0+\Omega_P-\Omega_R}\] ,
\end{eqnarray}
along with the analogous approximation for the negative energy
counterpart. The vector $P^{\mu}/p$ can finally be approximated by 
the unit vector $(1,\hat{p}^i)$ within the positive part and by 
$(-1,\hat{p}^i)$ within the negative one. This estimate is valid only 
at leading order and leaves soft subleading terms unconsidered. Since 
again soft corrections are unimportant, these vectors can be replaced 
respectively by $v^{\mu}$ and $-\bar{v}^{\mu}$ in order to satisfy 
strictly Ward identities for the vertex. The mechanism of algebraic 
cancellation between denominators and numerators with damping then occurs
\begin{eqnarray}
\tilde{V}^{1^{\mu}}_{RAR} & & (P,Q,-R)=\sum_{i=t,l}(-ie^2)
\int[dL]^i_{\rho\sigma}
\gamma^{\rho}\Ps \gamma^{\sigma}
\( \frac{v^{\mu}}{q_0+\Omega_{P+L}-\Omega_{R+L}}\[ 
\Delta^+_R(P+L)\right.\right.\nonumber\\&
&\left.\left.-\Delta^+_A(R+L)\]
+\frac{\bar{v}^{\mu}}{q_0+\Omega_{R+L}-\Omega_{P+L}}\[ \Delta^-_R(P+L)
-\Delta^-_A(R+L)\] \) . 
\end{eqnarray}
A basic point is to see to which extent the common denominators without
damping can be extracted from the integral. Taking explicitly the
expressions $\Omega_{P+L}$ and $\Omega_{R+L}$, namely $\sqrt{(\vec{p}
+\vec{l})^2}$ and $\sqrt{(\vec{r}+\vec{l})^2}$, with the soft terms from 
$L$ and $Q$, an expansion of these denominators can be written as
\begin{eqnarray}
\label{expansion}
q_0+\Omega_{P+L}-\Omega_{R+L} &=& q_0-\hat{p}.\vec{q}\( 1-
\frac{m_{\infty}^2}{2p^2}\) -\frac{q^2}{2p}(1-\hat{p}.\hat{q})
-\frac{\vec{q}}{p}.\[ \vec{l}
-\hat{p}(\hat{p}.\vec{l})\] \nonumber\\&
&+ q F(\vec{p},\vec{q},\vec{l})+O(e^4)\nonumber\\
&=& q_0+\Omega_P-\Omega_R -\frac{\vec{q}}{p}.\[ \vec{l}
-\hat{p}(\hat{p}.\vec{l})\] +O(e^3) 
\end{eqnarray}  
where $F$ is a function of the three vectors and of order $e^2$ at
most. In the infrared limit ($q_0,q\sim O(e^2T)$ and below) the terms
following $q_0+\Omega_P-\Omega_R$ can always be neglected. The
difference $q_0+\Omega_P-\Omega_R$ is equal to $q_0-\hat{p}\vec{q}$ and 
the asymptotic mass gives a subleading contribution (except on the
light-cone). On the light-cone if $\hat{p}.\hat{q}\sim\pm 1+O(e)$ the
subsequent terms remain of order $e^3T$ ($\hat{q}\sim\hat{p}$  and 
$(\vec{q}/p).( \vec{l}-\hat{p}(\hat{p}.\vec{l}))\sim O(e^3T)$) and can
be also discarded. However when the presence of the asymptotic mass
becomes necessary ($\hat{p}.\hat{q}\sim \pm 1+O(e^2)$), 
$q_0+\Omega_P-\Omega_R$ is of the same order as the other contributions 
and cannot be extracted. Therefore, except when 
$q_0+\Omega_P-\Omega_R\sim O(e^3T)$ on the light-cone, the replacements 
$1/(q_0+\Omega_{P+L}-\Omega_{R+L})$ by $1/(q_0+\Omega_P-\Omega_R)$ and 
$1/(q_0+\Omega_{R+L}-\Omega_{P+L})$ by $1/(q_0+\Omega_R-\Omega_P)$ 
are always justified and the denominators above may be extracted from 
the integral. The result for the 'one-loop' vertex related to the simple
ladder resummation is 
\begin{eqnarray}
\tilde{V}_{RAR}^{1^{\mu}}(P,Q,-R) &=& 
\frac{v^{\mu}}{q_0+\Omega_P-\Omega_R}\(
\Sigma^{1^+}_R(P)-\Sigma^{1^+}_A(R)\) \nonumber\\&
&\qquad\qquad+\frac{\bar{v}^{\mu}}{q_0+\Omega_R-\Omega_P}\( 
\Sigma^{1^-}_R(P)-\Sigma^{1^-}_A(R)\) ,
\end{eqnarray}
with the 'one-loop' self-energies with resummed fermion and photon 
propagators and the simplest vertex correction. Thus this first step 
clearly indicates that it is necessary to go beyond ladder resummation 
in order to recover the complete expression [\ref{completeform}] 
of the vertex.
\begin{figure}
\begin{center}
\leavevmode
\epsfxsize=4 in
\epsfbox{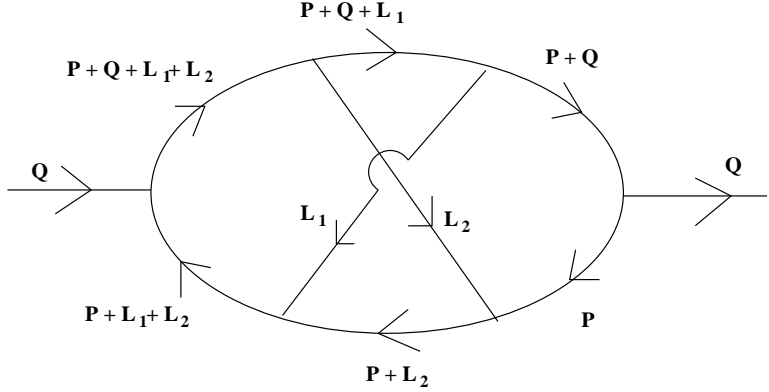}
\end{center}
\caption{A three--loop self--energy non--planar graph (with the same
conventions for the photon and electron lines as in Fig.~\ref{ladderfig}).}
\label{crossfig}\end{figure}
\subsection{Two-loop diagrams resummation}

In the previous subsection, only the simple ladder resummation has been
considered. It is now important to go beyond this and to take into account 
vertex corrections and crossed diagrams since these graphs have been
shown to contribute also at leading order. Therefore the 'two-loop'
graph of the vertex has to be investigated (see Fig.~\ref{crossfig}). 
Still fermion propagators are resummed with the 'complete' leading order
self-energy. The procedure leading to an algebraic cancellation of 
self-energies is much similar to the previous one with ladders.
\par
Considering the 'two-loop' crossed vertex when inserting the 
vertex [\ref{completeform}] gives
\begin{eqnarray}
\tilde{V} & & ^{2^{\mu}}_{Cr_{RAR}}(P,Q,-R) = e^4\int[dL_1]^t_{\alpha\beta}
[dL_2]^t_{\sigma\rho}\Delta_A(R+L_1)\gamma^{\alpha}(\Rs +\Los )
\Delta_A(R+L_1+L_2)\nonumber\\&
&\gamma^{\sigma}(\Rs +\Los +\Lts )\Delta_R(P+L_1+L_2)
\( \gamma^{\mu}+\tilde{V}^{\mu}_{RAR}(P+L_1+L_2,Q,-R-L_1-L_2)\)
\nonumber\\&
&(\Ps +\Los +\Lts )\gamma^{\beta}\Delta_R(P+L_2)(\Ps +\Lts )\gamma^{\rho}.
\end{eqnarray} 
The product of propagators $\Delta(P+L_1+L_2)\Delta(R+L_1+L_2)$ attached
to the internal vertex must be split in the same way as for the
ladders. Also the canonical decomposition of all the fermion propagators
into positive and negative energy parts enables to separate the positive
energy contribution of the expression above from its negative
counterpart. As it was pointed out before, mixed terms
(positive/negative energy terms) are shown to be subleading. The
algebraic cancellation of the self-energies still occurs and is
expressed through the replacement of  
$(q_0\pm\Omega_{P+L_1+L_2}\mp\Omega_{R+L_1+L_2}\pm\sigma_{R_{P+L_1+L_2}}
\mp\sigma_{A_{R+L_1+L_2}})$ by the denominator
$(q_0\pm\Omega_{P+L_1+L_2}\mp\Omega_{R+L_1+L_2})$. This yields
\begin{eqnarray}
\tilde{V}& &^{2^{\mu}}_{Cr_{RAR}}(P,Q ,-R) = ie^4\int[dL_1]^t_{\alpha\beta}
[dL_2]^t_{\sigma\rho}\Delta^+_A(R+L_1)\Delta^+_R(P+L_2)\gamma^{\alpha}
(\Rs +\Los ) \gamma^{\sigma}\nonumber\\&
&(\Rs +\Los +\Lts )(\Ps +\Los +\Lts )\gamma^{\beta}(\Ps +\Lts )\gamma^{\rho}
\frac{v^{\mu}}{q_0+\Omega_{P+L_1+L_2}-\Omega_{R+L_1+L_2}}\times\nonumber\\&
&\times\( \Delta^+_R(P+L_1+L_2)-\Delta^+_A(R+L_1+L_2)\) +n.e.
\end{eqnarray} 
where the negative energy part has the same expression as its positive
counterpart except for the propagators $\Delta^-$ and the denominator 
$\bar{v}^{\mu}/(q_0+\Omega_{R+L_1+L_2}-\Omega_{P+L_1+L_2})$. In the 
infrared limit and on the light-cone ($\hat{p}.\hat{q}\sim\pm 1+O(e)$), 
these denominators can be replaced by $(q_0+\Omega_P-\Omega_R)$ and 
$(q_0+\Omega_R-\Omega_P)$, since again the $L_i$-dependent terms may be 
neglected. It turns out
\begin{eqnarray}
\label{crossed}
\tilde{V}& &^{2^{\mu}}_{Cr_{RAR}}(P,Q,-R) = 
\frac{v^{\mu}}{q_0+\Omega_P-\Omega_R}ie^4\int[dL_1]^t_{\alpha\beta}
[dL_2]^t_{\sigma\rho}\Delta^+_A(R+L_1)\Delta^+_R(P+L_2)\nonumber\\&
&\gamma^{\alpha}(\Rs +\Los )\gamma^{\sigma}(\Rs +\Los +\Lts )
(\Ps +\Los +\Lts )\gamma^{\beta}(\Ps +\Lts )\gamma^{\rho}
\( \Delta^+_R(P+L_1+L_2)\right.\nonumber\\&
&\left. -\Delta^+_A(R+L_1+L_2)\) +n.e. 
\end{eqnarray} 
\par 
If now the 'two-loop' diagram with a vertex correction connected to the
line of momentum $P$ is considered, its expression reads
\begin{eqnarray}
\tilde{V}& &^{2^{\mu}}_{V1_{RAR}}(P,Q,-R) = e^4\int[dL_1]^t_{\alpha\beta}
[dL_2]^t_{\sigma\rho}\Delta_A(R+L_1)
\Delta_R(P+L_1)\Delta_R(P+L_1+L_2)\nonumber\\&
&\Delta_R(P+L_2)\gamma^{\alpha}(\Rs +\Los ) 
\( \gamma^{\mu}+\tilde{V}^{\mu}_{RAR}(P+L_1,Q,-R-L_1)\)
(\Ps +\Los )\gamma^{\sigma}\nonumber\\&
&(\Ps +\Los +\Lts )\gamma^{\beta}(\Ps +\Lts )\gamma^{\rho}.
\end{eqnarray} 
Again the product of propagators attached to the internal vertex,
in this case $\Delta(P+L_1)\Delta(R+L_1)$, has to be split. In the
very same way as before, the 'complete' self-energies get eliminated in
the common denominators. The infrared limit and the 'weak' light-cone
limit ($\hat{p}.\hat{q}\sim\pm 1+O(e)$ is the angle measuring the 
collinearity between the emitted photon and the fermion) allow to
neglect the $L_i$-dependent terms in these denominators. Therefore the 
latter can be approximated by $q_0+\Omega_P-\Omega_R$ and 
$q_0+\Omega_R-\Omega_P$, and finally extracted from the integral. The 
decomposition between positive energy and negative energy parts of the 
propagators has to be used and by discarding subleading pieces mixing 
positive and negative energy terms leads to two contributions. This gives
\begin{eqnarray}
\label{vertexun}
\tilde{V}& &^{2^{\mu}}_{V1_{RAR}}(P,Q,-R) = 
\frac{v^{\mu}}{q_0+\Omega_P-\Omega_R}ie^4\int[dL_1]^t_{\alpha\beta}
[dL_2]^t_{\sigma\rho}\Delta^+_A(P+L_1+L_2)\nonumber\\&
&\Delta^+_R(P+L_2)\gamma^{\alpha}
(\Ps +\Los )\gamma^{\sigma}(\Ps +\Los +\Lts )
(\Ps +\Los +\Lts )\gamma^{\beta}(\Ps +\Lts )\gamma^{\rho}\nonumber\\&
&\( \Delta^+_R(P+L_1)-\Delta^+_A(R+L_1)\) +n.e. 
\end{eqnarray} 
Finally the last 'two-loop' graph is the symmetric counterpart of the 
previous diagram with a vertex correction along the line
carried by the momentum $R$. Repeating the same procedure and splitting
this time the product $\Delta(P+L_2)\Delta(R+L_2)$, its expression in
the infrared limit and on the light-cone 
($\hat{p}.\hat{q}\sim\pm 1+O(e)$) becomes
\begin{eqnarray}
\label{vertexdeux}
\tilde{V}& &^{2^{\mu}}_{V2_{RAR}}(P,Q,-R) = 
\frac{v^{\mu}}{q_0+\Omega_P-\Omega_R}ie^4\int[dL_1]^t_{\alpha\beta}
[dL_2]^t_{\sigma\rho}\Delta^+_A(R+L_1)\nonumber\\&
&\Delta^+_A(R+L_1+L_2)\gamma^{\alpha}
(\Rs +\Los )\gamma^{\sigma}(\Rs +\Los +\Lts )
(\Ps +\Los +\Lts )\gamma^{\beta}(\Ps +\Lts )\gamma^{\rho}
\nonumber\\&
&\( \Delta^+_R(P+L_2)-\Delta^+_A(R+L_2)\) +n.e. 
\end{eqnarray} 
Therefore both, in the infrared limit and on the light-cone 
($\hat{p}.\hat{q}\sim\pm 1+O(e)$) the procedure is equivalent to 
a simple contraction with $Q^{\mu}$ times the vector 
$v^{\mu}/(q_0+\Omega_P-\Omega_R)$ (and its negative energy counterpart).
Adding the three previous 'two-loop' diagrams leads to the difference of
the self-energies at 'two-loop' order (with resummed fermion propagator
and the simplest vertex correction).
\begin{eqnarray}
\tilde{V}_{RAR}^{2^{\mu}}(P,Q,-R) &=& 
\frac{v^{\mu}}{q_0+\Omega_P-\Omega_R}\(
\Sigma^{2^+}_R(P)-\Sigma^{2^+}_A(R)\) \nonumber\\&
& \qquad +\frac{\bar{v}^{\mu}}{q_0+\Omega_R-\Omega_P}\( 
\Sigma^{2^-}_R(P)-\Sigma^{2^-}_A(R)\) .
\end{eqnarray} 
Since this corresponds to Ward identities applied to resummed diagrams
it is natural to repeat the same operation at any 'N-loop' order, 
{\it i.e.} with vertices which cannot be disconnected by cutting two
fermion lines.    
\subsection{N-loop diagrams resummation}

In the following 'N-loop vertices' are considered. The 'N-loop
photon-electron-electron vertices' are the generalization of the
previous ladder and 'two-loop graphs'. They involve resummed fermion
propagators and the exchange of soft transverse photons. They cannot 
be disconnected by cutting two fermion lines. 
It is shown that for any of such 'N-loop diagrams' the action of
the complete vertex of Eq.~[\ref{completeform}] is equivalent to the
contraction with the photon momentum $Q^{\mu}$. It is therefore natural
to rely on Ward identities at any order to prove that the
afore-mentioned vertex solves the Bethe-Salpeter equation in the
infrared limit.
\par
The 'N-loop graphs' are all related to a specific diagram of the
self-energy with vertex corrections via Ward identities and form a
subset of 'N-loop vertices'. Each 'N-loop diagram' can be written as 
\begin{eqnarray}
\tilde{V}& &^{N^{\mu}}_{M\, K_{RAR}}(P,Q,-R) = e^{2N}\int[dL_1]^t_{\alpha_1
\beta_1}
[dL_2]^t_{\alpha_2\beta_2}...[dL_N]^t_{\alpha_N\beta_N}\gamma^{\alpha_1}
(\Rs + \Los )\nonumber\\&
&\Delta_A(R+L_1)...(\Rs +\Lis +...+\Ljs )\Delta_A(R+L_i+...+L_j)
\Delta_R(P+L_i+...+L_j)
\nonumber\\&
&\( \gamma^{\mu}+\tilde{V}^{\mu}_{RAR}(P+L_i+...+L_j,Q,-R-L_i-...-L_j)\)
(\Ps +\Lis +...+\Ljs )...\nonumber\\&
\end{eqnarray}  
where $P+L_i+...+L_j$ and $R+L_i+...+L_j$ correspond to the internal
legs attached to the external photon line. From the arguments of the
previous sections the leading contribution comes from the tree-like
terms involving cut internal photon lines with the Bose-Einstein
factors. Each fermion propagator can be split along the
ways described before. That enables to split the vertex into a positive
energy  and a negative energy part. Also sticking the complete vertex of
Eq.~[\ref{completeform}] into $\tilde{V}^{N^{\mu}}_K$ leads to the usual
replacement of the common denominator $(q_0+\Omega_{P+L_i+...}-\Omega
_{R+L_i...}+\sigma_{R_{P+L_i+...}}-\sigma_{A_{R+L_i+...}})$ by $(q_0+\Omega_
{P+L_i+...}-\Omega_{R+L_i...})$. This follows stricly the lines of the
previous sections. Then in the infrared region and near the
light-cone without a too strong collinearity ($\hat{p}.\hat{q}\sim\pm
1+O(e)$ the $L_i$-dependent terms may be neglected. The same reasoning
can then be applied here as for the simple ladder resummation. It is enough
to replace $\vec{l}$ in the expansion of Eq.~[\ref{expansion}] by a more
general vector like $\vec{l_i}+...+\vec{l_j}$. Discarding negligible 
contributions involving terms such as $P^2$ gives
\begin{eqnarray}
\tilde{V} & & ^{N^{\mu}}_{M\, K_{RAR}}(P,Q,-R) =ie^{2N}\int[dL_1]^t_{\alpha_1
\beta_1}
[dL_2]^t_{\alpha_2\beta_2}...[dL_N]^t_{\alpha_N\beta_N}\gamma^{\alpha_1}
(\Rs + \Los )\nonumber\\&
&\Delta_A^+(R+L_1)...\frac{v^{\mu}}{q_0+\Omega_{P+L_i+...+L_j}
-\Omega_{R+L_i+...+L_j}}\( \Delta_R^+(P+L_i+...+L_j)\right. \nonumber\\&
&\left. -\Delta_A^+(R+L_i+...+L_j)\) (\Ps +\Lis +...+\Ljs )...+n.e.
\end{eqnarray}  
As explained above the common denominator can be extracted from the
integral in the same specific cases
\begin{eqnarray}
\tilde{V} & & ^{N^{\mu}}_{M\, K_{RAR}}(P,Q,-R) = 
\frac{v^{\mu}}{q_0+\Omega_P-\Omega_R}
ie^{2N}\int[dL_1]^t_{\alpha_1\beta_1}
[dL_2]^t_{\alpha_2\beta_2}...[dL_N]^t_{\alpha_N\beta_N}\gamma^{\alpha_1}
\nonumber\\&
&(\Rs + \Los )\Delta_A^+(R+L_1)...
\( \Delta_R^+(P+L_i+...+L_j)-\Delta_A^+(R+L_i+...+L_j)\) 
\nonumber\\&
&(\Ps +\Lis +...+\Ljs )...+n.e.
\end{eqnarray}
Therefore this is completely equivalent to a contraction of each graph
with $Q^{\mu}$ times the vector $v^{\mu}/(q_0+\Omega_P-\Omega_R)$ (and
its negative energy counterpart
$\bar{v}^{\mu}/(q_0+\Omega_R-\Omega_P))$. The expressions obtained 
correspond to specific parts, the sum of these parts giving after 
cancellations a particular self-energy diagram.
\begin{eqnarray}
\tilde{V}^{N^{\mu}}_{M\, K_{RAR}}(P,Q,-R) &=& \frac{v^{\mu}}
{q_0+\Omega_P-\Omega_R}
\( \Sigma^{N^+}_{M\, K_R}(P)-\Sigma^{N^+}_{M\, K_A}(R)\) \nonumber\\&
& \qquad +\frac{\bar{v}^{\mu}}{q_0+\Omega_R-\Omega_P}
\( \Sigma^{N^-}_{M\, K_R}(P)-\Sigma^{N^-}_{M\, K_A}(R)\) .
\end{eqnarray}  
The above procedure may be repeated for each graph. Adding all the
subset of diagrams (indice $K$) corresponding to a self-energy with 
a specific vertex correction 
\begin{equation}
\Sigma^{N^{\pm}}_{M_R}(P)=\sum_K\Sigma^{\pm}_{M\, K_R}(P)
\end{equation}
and adding each self-energy with a particular vertex correction (indice 
$M$) gives the 'N-loop' self-energy
\begin{equation}
\Sigma^{N^{\pm}}_R(P)=\sum_M\Sigma^{\pm}_{M_R}(P).
\end{equation}
Finally the sum of all the 'N-loop' self-energies, starting from the 
previous 'one-loop' (transverse and longitudinal photon exchanges) 
and 'two-loop' diagrams (only transverse) leads to the 'complete' 
self-energy at leading order. With the vector $v^{\mu}/(q_0+\Omega_P
-\Omega_R)$ (and $\bar{v}^{\mu}/(q_0+\Omega_R-\Omega_P)$ in front, the 
vertex of Eq.~[\ref{completeform}] is recovered and therefore shown 
to be the solution of the Bethe-Salpeter equation.    
\section{Cancellation of damping terms}

In this section the polarization tensor $\Pi^{\mu\nu}(Q)$ at leading
order ({\it i.e.} the order $e^2T^2$) is considered. It is worth
emphasizing again that this is no longer valid for subleading
quantities, especially $\Pi^{\mu}_{\mu}(Q)$. This is due to the  
approximations made when deriving the vertex of
Eq.~[\ref{completeform}]. In the infrared limit the latter is shown to
be the solution of the Bethe-Salpeter equation. In the case of the 
light-cone limit, the afore-mentioned vertex is not the complete
solution (in particular when the emission angle $\hat{p}.\hat{q}$
approaches $\pm 1$). It is nevertheless always interesting to look for the 
expression of $\Pi^{\mu\nu}(Q)$ provided by this vertex even in that
case. The main result is that with a resummation of the fermion
propagators with a damping, the latter drops out in the final
expression as it was pointed out in Ref.~\cite{smilga,kraemmer}. But 
unlike \cite{smilga,kraemmer} the asumption of a constant damping is not
required. Also here the general infrared limit (outside the light-cone)
has been investigated, not only specific terms such as $\Pi^{00}(q_0,0)$
and $\Pi^{ii}(0,q\rightarrow 0)$. Finally there are no ambiguities any
longer in denominators like $1/P.Q$ due to the absence of 
$i\epsilon$ prescriptions as it was the case with the vertex advocated
in Ref.~\cite{smilga}. 
\par
In the $R/A$ formalism, the retarded part of the tensor can be written as
\begin{eqnarray}
i\Pi^{\mu\nu}_{RR}(Q) &=& -e^2\Tr \int\frac{d^4P}{(2\pi)^4}\( \Ps 
\gamma^{\mu}\Rs \) 
\Bigg\{ \(
\frac{1}{2}-n_F(p_0)\) \Delta_R(R)\Big[ \(
\gamma^{\nu}\right.\nonumber\\&
&\left.\left. +\tilde{V}^{\nu}_{RRA}
(P,Q,-R)\) \Delta_R(P)-\( \gamma^{\nu}+\tilde{V}^{\nu}_{ARA}(P,Q,-R)\) 
\Delta_A(P)\] \nonumber\\&
&+\( \frac{1}{2}-n_F(r_0)\) \Delta_A(P)
\[ \( \gamma^{\nu}+\tilde{V}^{\nu}_{ARA}(P,Q,-R)\) 
\Delta_R(R)\right.\nonumber\\&
&\left.-\( \gamma^{\nu}+
\tilde{V}^{\nu}_{ARR}(P,Q,-R)\) \Delta_A(R)\] \Bigg\} .
\end{eqnarray}
Contracting the spinors gives a part sensitive to the infrared or
light-cone region plus possible tadpole terms. These tadpoles are not 
concerned by vertex and damping corrections. Taking $\Pi^{00}$ as a
particular example, the relevant expression is 
\begin{eqnarray}
i\Pi & &^{00}_{RR}(Q)= -e^2\int\frac{d^4P}{(2\pi)^4}8p_0^2
\left\{ \(
\frac{1}{2}-n_F(p_0)\) \[ \Bigg( 1-\frac{\sigma^+_R(P)
-\sigma^+_R(R)}{q_0+\Omega_P-\Omega_R}-\right.\right.
\nonumber\\&
&\left.\frac{\sigma^-_R(P)-\sigma^-_R(R)}
{q_0+\Omega_R-\Omega_P}\) \Delta_R(R)\Delta_R(P) 
-\Delta_R(R)\Delta_A(P)\( 1-\frac{\sigma^+_A(P)
-\sigma^+_R(R)}{q_0+\Omega_P-\Omega_R}-\right.\nonumber\\&
&\left.\left.\left.\frac{\sigma^-_A(P)-\sigma^-_R(R)}
{q_0+\Omega_R-\Omega_P}\) \] +\( \frac{1}{2}-n_F(r_0)\) \[ 
\Delta_R(R)\Delta_A(P)\( 1-\frac{\sigma^+_A(P)
-\sigma^+_R(R)}{q_0+\Omega_P-\Omega_R}-\right.\right.\right.\nonumber\\&
&\left.\left.\left.\frac{\sigma^-_A(P)-\sigma^-_R(R)}
{q_0+\Omega_R-\Omega_P}\) 
-\Delta_A(R)\Delta_A(P)\( 1-\frac{\sigma^+_A(P)
-\sigma^+_A(R)}{q_0+\Omega_P-\Omega_R}-\frac{\sigma^-_A(P)-\sigma^-_A(R)}
{q_0+\Omega_R-\Omega_P}\) \] \right\} .\nonumber\\&
\end{eqnarray}
Splitting the products of propagators give denominators
containing the retarded and advanced $\sigma$'s. These denominators
cancel against the numerators of the internal vertices written
above. What remains as usual are the differences between 
$\Delta^{\pm}(P)$ and $\Delta^{\pm}(R)$. More explicitly
\begin{eqnarray}
i\Pi & &^{00}_{RR}(Q)= -4ie^2\int\frac{d^4P}{(2\pi)^4}p\left\{ \(
\frac{1}{2}-n_F(p_0)\) \[ \frac{1}{q_0+\Omega_P-\Omega_R}\( \Delta^+_R(P)
-\Delta^+_R(R)\) \right.\right.\nonumber\\&
&\left.\left.\qquad -\frac{1}{q_0+\Omega_R-\Omega_P}\( \Delta^-_R(P)
-\Delta^-_R(R)\) -\frac{1}{q_0+\Omega_P-\Omega_R}\( \Delta^+_A(P)
-\Delta^+_R(R)\) \right.\right.\nonumber\\&
&\left.\left.+\frac{1}{q_0+\Omega_R-\Omega_P}\( \Delta^+_A(P)
-\Delta^+_R(R)\) \] +\( \frac{1}{2}-n_F(r_0)\) \[ \frac{1}{q_0+\Omega_P
-\Omega_R}\( \Delta^+_A(P) \right.\right.\right.\nonumber\\&
&\left.\left.\left.-\Delta^+_R(R)\) -\frac{1}{q_0+\Omega_R-\Omega_P}\(
\Delta^-_A(P)-\Delta^-_R(R)\) -\frac{1}{q_0+\Omega_P-\Omega_R}\( \Delta^+_A(P)
\right.\right.\right.\nonumber\\&
&\left.\left.\left.\qquad-\Delta^+_A(R)\) +\frac{1}{q_0+\Omega_R-\Omega_P}\(
\Delta^-_A(P)-\Delta^-_A(R)\) \] \right\} .         
\end{eqnarray}
It is then possible to get the cancellation of propagators, the $R/A$ 
prescriptions of which remain unchanged for the same statistical factor. 
This gives just cut propagators for the variable associated to the
statistical factor
\begin{eqnarray}
i\Pi & &^{00}_{RR}(Q)= -4ie^2\int\frac{d^4P}{(2\pi)^4}p\left\{ \(
\frac{1}{2}-n_F(p_0)\) \[ \frac{1}{q_0+\Omega_P-\Omega_R}\( \Delta^+_R(P)
-\Delta^+_A(P)\) \right.\right.\nonumber\\&
&\left.\left.-\frac{1}{q_0+\Omega_R-\Omega_P}\( \Delta^-_R(P)
-\Delta^-_A(P)\) \] -\( \frac{1}{2}-n_F(r_0)\) \[ \frac{1}{q_0+\Omega_P
-\Omega_R}\( \Delta^+_R(R) \right.\right.\right.\nonumber\\&
&\left.\left.\left.-\Delta^+_A(R)\) -\frac{1}{q_0+\Omega_R-\Omega_P}\(
\Delta^-_R(R)-\Delta^-_A(R)\) \] \right\} .         
\end{eqnarray}
It can be noticed at this stage that the {\it improved hard thermal
loop} is recovered if the dampings (imaginary parts of the $\sigma$'s) 
inside the propagators are replaced by the usual $i\epsilon$
prescriptions. In this case the differences of propagators just give
Dirac functions. At first glance this approximation by $\delta$
functions seems justified if the scale $e^2T$ of the 
damping compared to the hard scale is taken into account. However the 
expression above 
contains Breit-Wigner functions but with energy dependent widths. 
Thus it seems more appropriate to look for a rigourous treatment
of these terms. The main point is that the cancellation of
retarded-advanced products replaced by simple propagators allows to
convert the integral over $p_0$ into a contour integral and to choose 
the complex half-plane without the discontinuities from the damping
terms. For the retarded propagators of the expression above, a closed
contour in the upper half-plane avoiding the fermion Matsubara
frequencies $\omega_n=2i\pi(n+\frac{1}{2})T$ on the imaginary axis 
gives no contribution. This contour can be composed of the real axis,
the sum $C_1$ of two quarter-circles  expanding at infinity and a part
$C_2$ encircling clockwise the upper poles of the statistical
factor. The integration over the real axis can be replaced by an
integration over $-C_1-C_2$. Relabeling the variables $p_0$ and $r_0$ 
as $z$, the expression of the polarization tensor coming from the retarded
propagators is
\begin{eqnarray}
i\Pi^{00^{up}}_{RR} & & (Q)=
2e^2\int_{-C_1-C_2}\frac{dz}{2\pi}\int\frac{d^3p}{(2\pi)^3}
\left\{ \( \frac{1}{2}-n_F(z)\) \[
\frac{1}{q_0+\Omega_P-\Omega_R}\times
\right.\right.\nonumber\\&
&\left.\left.\times\( \frac{1}{z-\Omega_P+\sigma^+(z,p)}
-\frac{1}{z-\Omega_R+\sigma^+(z,r)}\) 
-\frac{1}{q_0+\Omega_R-\Omega_P}\times\right.\right.\nonumber\\&
&\left.\left.\times\( \frac{1}{z+\Omega_P+\sigma^-(z,p)}-
\frac{1}{z+\Omega_R+\sigma^-(z,r)}\) \] \right\} .\nonumber\\&         
\end{eqnarray} 
Each propagator taken individually yields a finite contribution
over $-C_1$. In order to get denominators falling off as $1/z^2$ when
$|z|$ tends to infinity, the function
\begin{equation}
f_{\pm}(z) = \( \frac{1}{2}-n_F(z)\) \( 
\frac{1}{z-\Omega_P+\sigma^{\pm}(z,p)}-\frac{1}{z-\Omega_R+\sigma^{\pm}
(z,r)}\) ,
\end{equation} 
must be considered. It has the property $zf_{\pm}(z)\rightarrow 0$ 
when $|z|\rightarrow\infty$, wherever it is analytical, which is the
case on $C_1$. Therefore it is a specific sum of propagators which gives
no contribution over $-C_1$. The next step consists in writting 
the contribution of $-C_2$, namely the sum of the residues of the 
Fermi-Dirac factor for the poles located in the upper half-plane. The
same reasoning can be applied for the advanced propagators with a
contour in the lower half-plane. The contribution is finally reduced to
the sum of the residues corresponding to the Matsubara frequencies in
the lower half-plane. Adding the parts from the retarded and advanced
propagators gives
\begin{eqnarray}
\label{matsuform}
i\Pi^{00}_{RR} & & (Q)=
4i\pi T e^2\int\frac{d^3p}{(2\pi)^3}\sum_n
  \[ \frac{1}{q_0+\Omega_P-\Omega_R}\: 
\frac{1}{\omega_n-\Omega_P+\sigma^+(\omega_n,p)}
\right.\nonumber\\&
&\left.-\frac{1}{q_0+\Omega_R-\Omega_P}
\frac{1}{\omega_n+\Omega_P+\sigma^-(\omega_n,p)} 
-\frac{1}{q_0+\Omega_P-\Omega_R}\:\frac{1}{\omega_n-\Omega_R+
\sigma^+(\omega_n,r)}\right.\nonumber\\&
&\left.+\frac{1}{q_0+\Omega_R-\Omega_P}
\frac{1}{\omega_n+\Omega_R+\sigma^-(\omega_n,r)} \] .    
\end{eqnarray}
The separation between the different scales can now be used to prove
the irrelevance of the damping contribution at leading order. The
frequencies $\omega_n$ all belong to the hard scale, along with
$\Omega_P$ or $\Omega_R$. The imaginary parts given by $\omega_n$ like 
the real part given by $\Omega_P$ or $\Omega_R$ overwhelm the real and 
imaginary parts of the $\sigma$'s. But the soft term $\Omega_P-\Omega_R
\sim\hat{p}.\vec{q}$ should intervene in the differences written
above. In the light-cone limit with $q\sim O(eT)$, the vertex and the
damping resummations are only relevant when the emission angle is close 
to $\pm 1$. In that case $\hat{p}.\vec{q}$ remains soft and much larger
than the $\sigma$'s. In the infrared case $q\sim O(e^2T)$, and 
$\hat{p}.\vec{q}$ has therefore the same order (for the main part of the
phase-space the angle is not too small and does not alter this
estimate). But the order of magnitude of the $\sigma$'s when going away
from the real axis should not exceed the scale of order $e^3T$ contrary
to what happens when $p_0\simeq\Omega_P$. A simple power counting shows 
that a simplified expression corresponding to the 'one-loop' self-energy 
is $e$ times the order of $\hat{p}.\vec{q}$. Therefore in all these
cases, provided that the value of $q$ is not too small (below
$O(e^2T)$), the self-energies contributions should be negligible.   
What is left is nothing else than the term obtained with the simple 
$i\epsilon$ prescriptions and delta functions in real time. Neglecting 
the $\sigma$'s, the usual procedure of deforming the contours when going
from the imaginary time to the real time formalism can be used. The 
contribution given by simple poles is re-established
\begin{eqnarray}
i\Pi & &^{00}_{RR}(Q)= -4i\pi e^2\int\frac{d^4P}{(2\pi)^4}
\left\{ \(
\frac{1}{2}-n_F(p_0)\) \[ \frac{1}{q_0+\Omega_P-\Omega_R}
\delta(p_0-\Omega_P)\right.\right.\nonumber\\&
&\left. -\frac{1}{q_0+\Omega_R-\Omega_P}
\delta(p_0+\Omega_P)\Bigg] 
-\( \frac{1}{2}-n_F(r_0)\) \[ \frac{1}{q_0+\Omega_P
-\Omega_R}\delta(r_0-\Omega_R)\right.\right.\nonumber\\&
&\left.\left. -\frac{1}{q_0+\Omega_R-\Omega_P)}
\delta(r_0+\Omega_R)\] \right\} ,  
\end{eqnarray}
and the {\it improved hard thermal loop} is recovered. In the form
written above, it is a complex $q_0=\Re\,q_0+i\epsilon$ ($-i\epsilon$ 
for the advanced Green function) which is considered. This is
explicitly mentioned in the previous sections. This allows to obtain
straightforwardly the imaginary part or Landau damping
contribution. In the estimate above, it can be seen from 
Eq.~[\ref{matsuform}] that the limits $\Pi^{00}(q_0,0)$ or  
$\Pi^{ii}(0,q\rightarrow 0)$ have a particular status since 
$\Omega_R\rightarrow\Omega_P$ and 
$\sigma(\omega_n,r)\rightarrow\sigma(\omega_n,p)$. They both involve
terms such as $(\Omega_P-\Omega_R+\sigma(\omega_n,p)-\sigma(\omega_n,r))
/(\omega_n-\Omega_P+\sigma(\omega_n,p))^2$ which can be approximated by 
$(\Omega_P-\Omega_R)/(\omega_n-\Omega_P)^2$ for any value of $q$ below
the order $e^2T$. This leads to the same answers as without damping
contributions. 
\par
In short, sticking the expression of the complete vertex into
the polarization tensor at leading order gives the {\it improved 
hard thermal loop} introduced in Ref.~\cite{flech}. However this
does not give the complete solution in the light-cone limit. In the 
infrared limit, where the vertex satisfying the Bethe-Salpeter equation
can be found, the tensor just corresponds to the usual {\it hard thermal
loop} as expected.
\section{Conclusion}

It has been shown that in the infrared and in the light-cone region in 
$QED$ ladders are not the only leading diagrams for the resummed vertex when 
going beyond a simplified model. Due to long range magnetic interactions, 
specific non-planar graphs and graphs with vertex corections have to be 
considered at leading order. They all involve exchanges of soft photons 
and resummed fermion propagators. Both, in the infrared limit and in a weak 
light-cone limit an improved vertex could be derived which solves the 
Bethe-Salpeter equation. The resummation of all these 
vertex diagrams cancels against self-energy insertions containing soft 
photons. This compensation is algebraic for the most part. 
Technical ambiguities are lifted when the products of propagators with
different $R/A$ prescriptions cancel out. This doe not require any 
particular input on the self-energies, but only very general
properties. The result 
is found to be the {\it improved hard thermal loop} expression, which in the 
infrared or weak light-cone limits just corresponds to the usual 
$HTL$ term. In the strong light-cone limit where the effects of the asymptotic 
mass become important, the improved vertex is no longer a solution of the 
Bethe-Salpeter equation. Basic approximations valid in the former cases are 
no longer legitimate. Other methods for solving this problem are necessary and 
will be the subject of future work \cite{pet}. 
%
\clearpage
\appendix{APPENDIX}
\section{Two-loop self-energy}

In this appendix the two-loop diagram of the self-energy is calculated 
using the correct transverse spectral density $\rho_T$ (more exactly
its Landau damping part). The purpose is clearly to consider all the 
relevant scales for the internal momenta, and not only the very soft 
regime $O(e^2T)$. It is also interesting to see how a complete 
calculation which includes dynamical screening reaches the 
conclusion already established with the approximate spectral density 
given in Eq.~[\ref{approspectral}]. 
\par
The starting point is given by the the form of Eq.~[\ref{zetaplus}], 
where $p_0=p$. After recombining the denominators by change of
variables, this yields  
\begin{eqnarray}
\zeta(P)& =& \frac{1}{4p_0}\Tr \( \Ps \Sigma_{RR}(P)\) 
  \nonumber\\&  
  =&(e^2T)^2\int\frac{d^3q}{(2\pi)^3}\int\frac{dq_0}{2\pi q_0}\rho_T(Q)
      \int\frac{d^3k}{(2\pi)^3}\int\frac{dk_0}{2\pi k_0}\rho_T(K)
      \( 1-(\hat{p}\cdot\hat{q})^2\) \nonumber\\&
 &\( 1-(\hat{p}\cdot\hat{k})^2\) \( i\pi\delta(q_0+k_0-\hat{p}\cdot
(\vec{q}+\vec{k}))\frac{\pp}{(q_0-\hat{p}\cdot\vec{q})^2} \right.
\nonumber\\&
&\left. \qquad\qquad\qquad\qquad\qquad\qquad
 -2i\pi\delta(k_0-\hat{p}\cdot\vec{k})
 \frac{\pp}{(q_0-\hat{p}\cdot\vec{q})^2}\) .
\end{eqnarray}
This expression may be split into two parts, associated respectively to 
$\delta(k_0-\hat{p}\cdot\vec{k})$ and
$\delta(q_0+k_0-\hat{p}\cdot(\vec{q}+\vec{k}))$. The first part can 
easily be computed with the correct spectral densities, due to a complete
separation (or factorization) between the terms involving $Q$ and $K$. 
First the integration over the angles gives
\begin{eqnarray}
\zeta_1(P)& =& 2i\pi (e^2T)^2 \int_{\mu}^{\infty}\frac{dk}{(2\pi)^2}k
      \int_{-k}^k\frac{dk_0}{2\pi k_0}\rho_T(K)\( 1-\frac{k_0^2}{k^2}\) 
\nonumber\\&
   &\qquad\qquad\qquad\qquad\int_{\mu}^{\infty}\frac{dq}{(2\pi)^2}
     \int_{-q}^q\frac{dq_0}{2\pi}
   \rho_T(Q)\( \frac{4}{q_0}+\frac{2}{q}\ln\left|
      \frac{q_0-q}{q_0+q}
\right| \) . 
\end{eqnarray}
The well-known sum rules \cite{braay,piz} read:
\begin{equation}
\label{sumrul}
\int_{-k}^k\frac{dk_0}{2\pi k_0}\rho_T(K)= \frac{1}{k^2}-\frac
{Z_T(k)}{\omega_T^2(k)}, \qquad
\int_{-k}^k\frac{dk_0}{2\pi}k_0\rho_T(K)= 1-Z_T(k),
\end{equation}
where $Z_T(k)$ and $\omega_T(k)$ are the residue and the transverse 
dispersion relation, respectively. These sum rules can be used for
carrying out the integration over $q_0$. However, the two-loop graph is 
only at leading order, when $q$ can reach the infrared limit, {\it i.e.} 
the order $e^2T$. Thus the logarithmic term is shown to be negligible. 
Since there is clearly a separation between two scales, the soft one of 
order $eT$ and the infrared or very soft scale of order $e^2T$, an
intermediate scale parameter $k^*\sim O(e\sqrt{e}T)$ can be introduced. 
The first part may then be written as
\begin{eqnarray}
\label{finalsep}
\zeta_1(P) & =& \frac{4i}{(2\pi)^3}(e^2T)^2\frac{1}{\mu}
   \int_{\mu}^{\infty}\frac{dk}{k}Z_T(k)\( 1-\frac{k^2}{\omega_T^2(k)}\)
   \nonumber\\&
 =& \frac{4i}{(2\pi)^3}(e^2T)^2\frac{1}{\mu}\ln\( \frac{k^*}{\mu}\) 
  +\frac{4i}{(2\pi)^3}(e^2T)^2\frac{1}{\mu}
   \int_{k^*}^{\infty}\frac{dk}{k}Z_T(k)\( 1-\frac{k^2}{\omega_T^2(k)}\) ,
\end{eqnarray}
where the integral has been decomposed into an infrared and a soft
scale contribution.
\par
The calculation of the second part is by far more complicated, due to 
the non trivial phase-space. The kinematical constraints give, for
$k>q$:
\begin{eqnarray*}
1>\cos\theta >-1,  \qquad\qquad\qquad\qquad k-q > q_0+k_0>q-k, \qquad (1)\\
1> \cos\theta >\frac{q_0+k_0-k}{q}, \qquad\qquad\qquad q+k > q_0+k_0 > k-q, 
\qquad (2)\\
\frac{q_0+k_0+k}{q} > \cos\theta >-1, \qquad\qquad\qquad q-k > q_0+k_0> -q-k, \qquad (3)  
\end{eqnarray*}
and for $k<q$:
\begin{eqnarray*}
\frac{q_0+k_0+k}{q}>\cos\theta >\frac{q_0+k_0-k}{q},  \qquad\qquad 
q-k > q_0+k_0>k-q, \qquad (1')\\
1> \cos\theta >\frac{q_0+k_0-k}{q}, \qquad\qquad\qquad q+k > q_0+k_0 > q-k, 
\qquad (2')\\
\frac{q_0+k_0+k}{q} > \cos\theta >-1, \qquad\qquad\qquad k-q > q_0+k_0 >-q-k, 
\qquad (3')  
\end{eqnarray*}
Using the sum rules of Eq.~[\ref{sumrul}] seems useless with the
structure of the phase space. However under specific conditions 
further simplifications can still be made. First the case where
either $k$ or $q$ is not in the infrared limit, but still remains of
order $eT$ can be treated. The other variable must be necessarily
of order $e^2T$ in order to keep the two loop diagram dominant. The case
with both variables in the infrared region will be considered
afterwards. In
order to keep $k$ or $q$ of order $eT$, the scale parameter $k^*$
previously introduced may be used. With $k$ larger than $k^*$ for
instance, $q$ lies in the infrared. For very soft momenta, the whole 
density of states is concentrated around zero energy, $q_0\ll q$ and for
the dominant part $q\ll k^*\ll k$. With these conditions, the regions 
$(2)$ and $(3)$ give negligible contributions and region $(1)$ is
reduced to $-k<k_0<k$. The variable $q_0$ is still limited by positive
and negative values. The expression corresponding to this part can be 
written as:
\begin{eqnarray}
\zeta^{(1)}_2(P)& =& i\pi (e^2T)^2 \int_{k^*}^{\infty}\frac{dk}{(2\pi)^2}k
      \int_{-k}^k\frac{dk_0}{2\pi k_0}\rho_T(K) 
   \int_{\mu}^{k}\frac{dq}{(2\pi)^2}q^2\int\frac{dq_0}{2\pi
      q_0}\rho_T(Q)\nonumber\\&
  &\qquad\qquad
 \int_{-1}^1d\cos\theta\frac{(1-\cos^2\theta)}{(q_0-q\cos\theta)^2}
  \( 1-\frac{1}{k^2}(q_0+k_0-q\cos\theta)^2\) . 
\end{eqnarray}
Again keeping only dominant terms, with the strict condition $q_0\ll q$,
a simplified term for the trace in the numerator is obtained
\begin{equation}
\zeta^{(1)}_2(P) = \frac{-2i}{(2\pi)^3}(e^2T)^2
   \int_{k^*}^{\infty}\frac{dk}{k}Z_T(k)\( 1-\frac{k^2}{\omega_T^2(k)}\)
   \int_{\mu}^{k}dq\( \frac{1}{q^2}+\frac{1}{3k^2}\) . 
\end{equation}
The remaining leading terms are the divergent part $1/\mu$ and  $1/k^*$. This 
last contribution is expected to get canceled afterwards. Thus
\begin{equation}
\label{finalun}
\zeta^{(1)}_2(P) = \frac{-2i}{(2\pi)^3}(e^2T)^2\frac{1}{\mu}
   \int_{k^*}^{\infty}\frac{dk}{k}Z_T(k)\( 1-\frac{k^2}{\omega_T^2(k)}\)
   +\frac{4i}{(2\pi)^3}(e^2T)^2\frac{1}{3k^*} . 
\end{equation}
\par
With $q$ larger than $k^*$, $k$ is forced to lie in the infrared and 
the case $k_0\ll k$ has to be considered with the divergent contribution
given by $k\ll k^*\ll q$. Now regions $(2')$ and $(3')$ are negligible 
and region $(1')$ simplifies to $-q<q_0<q$ with the same
restrictions for the angle. The starting formula is then
\begin{eqnarray}
\zeta^{(1')}_2(P)& =& i\pi (e^2T)^2 \int_{\mu}^{k^*}\frac{dk}{(2\pi)^2}k
      \int\frac{dk_0}{2\pi k_0}\rho_T(K) 
   \int_{k^*}^{\infty}\frac{dq}{(2\pi)^2}q^2\int_{-q}^q\frac{dq_0}{2\pi
      q_0}\rho_T(Q)\nonumber\\&
  &\qquad\qquad
 \int_{\frac{q_0+k_0-k}{q}}^{\frac{q_0+k_0+k}{q}}d\cos\theta
 \frac{(1-\cos^2\theta)}{(q_0-q\cos\theta)^2}
  \( 1-\frac{1}{k^2}(q_0+k_0-q\cos\theta)^2\) . 
\end{eqnarray}
The use of similar simplifications gives
\begin{equation}
\label{finalunpr}
\zeta^{(1')}_2(P) = \frac{-2i}{(2\pi)^3}(e^2T)^2
   \int_{k^*}^{\infty}\frac{dq}{q}Z_T(q)\( 1-\frac{q^2}{\omega_T^2(q)}\)
   \int_{\mu}^{q}dk\( \frac{1}{k^2}+\frac{1}{3q^2}\) . 
\end{equation}
Inverting the names of the variables $K$ and $Q$ gives exactly the same 
contribution as Eq.~[\ref{finalun}]. The sum of the two divergent parts   
Eq.~[\ref{finalun}] and Eq.~[\ref{finalunpr}] exactly cancel against 
the term of Eq.~[\ref{finalsep}] where $k$ is restricted to the simple 
soft scale $k^*\ll k$. 
\par
Finally the contributions where both variables $k$ and $q$ lies in the 
infrared remain to be inspected. In these cases, simplifications concerning 
all the regions of the phase space can no longer be made. For the region
$(1)$ of the phase space, the integration over the angles leads to
\begin{eqnarray}
\label{completun}
\zeta^{(1)}_2(P)& =& -4i\pi (e^2T)^2 \int_{\mu}^{k^*}\frac{dq}{(2\pi)^2}q^2
      \int\frac{dq_0}{2\pi q_0}\rho_T(Q) 
   \int_{q}^{k^*}\frac{dk}{(2\pi)^2}k\nonumber\\&
  &\qquad\qquad\qquad\qquad\int_{q-k-q_0}^{k-q-q_0}dk_0
  \frac{3\omega_p^2}{4k}\frac{1}
  {k^4+\( \frac{3\pi \omega_p^2 k_0}{4k}\) ^2}\( \frac{1}{q^2}+
  \frac{1}{3k^2}\) .
\end{eqnarray} 
The Landau damping part of the spectral
density is taken for a very soft momentum $k\ll eT$. With the
dominant contribution, $k_0/k$ remains of order $e^2$. The same  
argument concerns also $q_0/q$ since $q$ is in the infrared, too. 
Therefore all the terms involving $q_0$ and $k_0$ in the expression 
of the trace in the numerator can reasonnably be neglected. But the 
point is to keep nevertheless $q_0$ in the bounds of integration over 
$k_0$, since it is not yet known under what circumstances $k-q$ becomes 
very small and comparable to $q_0$. Carrying out the integration over 
$k_0$ gives
\begin{eqnarray}
\zeta^{(1)}_2(P)& =& -4i\pi (e^2T)^2 \int_{\mu}^{k^*}\frac{dq}{(2\pi)^2}q^2 
   \int_{q}^{k^*}\frac{dk}{(2\pi)^2 k}\( \frac{1}{q^2}+\frac{1}{3k^2}\) 
   \int_{-\infty}^{\infty}\frac{dq_0}{2\pi}\frac{3\omega_p^2}{2q}
  \frac{1}{q^4+\( \frac{3\pi \omega_p^2 q_0}{4q}\) ^2}\nonumber\\& 
  &\( \arctan\( \frac{3\pi\omega_p^2(k-q-q_0)}{4k^3}\) + 
  \arctan\( \frac{3\pi\omega_p^2(k-q+q_0)}{4k^3}\) \) ,
\end{eqnarray} 
where the bounds for the integration over $q_0$ have been replaced by
$(-\infty,\infty)$ in order to make the next step of the calculation
tractable. This is allowed since, although the function is not
correct when $q_0/q$ is no longer of order $e^2$, the whole contribution
becomes subleading. At this stage the Parseval relation between
functions and their Fourier transforms may be used to perform the
integration over $q_0$. Thus, a two-dimensional integral is obtained
\begin{eqnarray}
\label{intermun}
\zeta^{(1)}_2(P) &=& -4i\pi (e^2T)^2 \int_{\mu}^{k^*}\frac{dq}{(2\pi)^2}
   \int_{q}^{k^*}\frac{dk}{(2\pi)^2 k}\( \frac{1}{q^2}+\frac{1}{3k^2}\) 
\nonumber\\&
&\qquad\qquad\qquad\qquad\qquad\qquad
   \frac{2}{\pi}\arctan\( \frac{3\pi\omega_p^2(k-q)}{4(k^3+q^3)}\) .
\end{eqnarray}
The integrations over $q$ and $k$ can be performed. Two cases we have 
to be considered when integrating over $q$ for instance. First the region 
where $q$ is close enough to $k$ to make  the $\arctan$ not equal to
$\pi/2$. Second the region where the $\arctan$ is actually equal to $\pi/2$
plus a negligible correction of $O(e)$. 
In both cases the replacement  
\begin{equation}
\arctan\( \frac{3\pi\omega_p^2(k-q)}{4(k^3+q^3)}\) \qquad\longrightarrow 
\qquad\arctan\( \frac{3\pi\omega_p^2(k-q)}{8k^3}\) ,
\end{equation}
renders the integration tractable. Without entering into details, it can
be shown that the leading term gives the same result as when directly 
replacing the $\arctan$ by $\pi/2$ in Eq.~[\ref{intermun}]. This 
result could be anticipated at the level of this equation, since the
region where the difference $k-q$ is very small gives a negligible
contribution, since no such terms appear in the denominator.    
The result for this contribution may be written as
\begin{equation}
\label{finalinfrarun}
\zeta^{(1)}_2(P) = \frac{-2i}{(2\pi)^3}(e^2T)^2\frac{1}{\mu}\( \ln\(
\frac{k^*}{\mu}\) -\frac{5}{6}\) -\frac{4i}{(2\pi)^3}(e^2T)^2\frac{1}{3k^*} . 
\end{equation}
As for the region $(1')$, the specific bounds of integration over the 
angle yields a different trace in the numerator compared to the previous
case. Once again terms proportional to the energies $k_0$ and $q_0$ 
can be neglected. The analogous formula of Eq.~[\ref{completun}] is
therefore   
\begin{eqnarray}
\zeta^{(1')}_2(P)& =& -4i\pi (e^2T)^2 \int_{\mu}^{k^*}\frac{dk}{(2\pi)^2}k^2
      \int\frac{dk_0}{2\pi k_0}\rho_T(K)\int_{k}^{k^*}\frac{dq}{(2\pi)^2}q
      \( \frac{1}{k^2}+\frac{1}{3q^2}\) \nonumber\\&
  &\qquad\qquad\qquad\qquad\int_{k-q-k_0}^{q-k-k_0}dq_0
  \frac{3\omega_p^2}{4q}\frac{1}
  {q^4+\( \frac{3\pi \omega_p^2 q_0}{4q}\) ^2},
\end{eqnarray}
Finally exchanging $Q$ and $K$ gives straightforwardly the same final 
answer as Eq.~[\ref{finalinfrarun}]
\begin{equation}
\label{finalinfrarunpr}
\zeta^{(1')}_2(P) = \frac{-2i}{(2\pi)^3}(e^2T)^2\frac{1}{\mu}\( \ln\(
\frac{k^*}{\mu}\) -\frac{5}{6}\) -\frac{4i}{(2\pi)^3}(e^2T)^2\frac{1}{3k^*}. 
\end{equation}
The final step concerns the contributions corresponding to the regions
$(2)$, $(3)$, $(2')$ and $(3')$. Focusing only on $\zeta^{(2)}_2$ and 
$\zeta^{(3)}_2$ new constraints must be taken into account (see the
inequalities of regions $(2)$ and $(3)$) for the trace in the
numerator. With these constraints, after neglecting subleading terms
and carrying out integrations over the energies in the same way as
before, it turns out 
\begin{eqnarray}
|\zeta^{(2)}_2(P) + \zeta^{(3)}_2(P)|& <& \frac{\pi}{(2\pi)^4}(e^2T)^2
\int_{\mu}^{k^*}dq\int_{q}^{k^*}\frac{dk}{k}
  \( \frac{3}{q^2}+\frac{2}{k^2}+\frac{1}{kq}\) \nonumber\\&
  &\frac{2}{\pi}\( \arctan\( \frac{3\pi\omega_p^2(k+q)}{4(k^3+q^3)}\) 
 - \arctan\( \frac{3\pi\omega_p^2(k-q)}{4(k^3+q^3)}\) \) .
\end{eqnarray}
The first $\arctan$ can obviously be replaced by $\pi/2$, since the two
momenta are in the infrared. The correction is always smaller by at
least a factor $O(e)$. For the second $\arctan$, the calculation can be
carried out in the same way as for the parts corresponding to $(1)$ and 
$(1')$. Again the region where $k-q$ is small gives a negligible 
contribution . Discarding corrections smaller by a factor $O(e)$, and 
keeping only the terms $(1/\mu)\ln(k^*/\mu)$, $1/\mu$ and $1/k^*$, 
this part cancels against its counterpart from the first $\arctan$. 
Therefore $\zeta^{(2)}_2(P)$ and $\zeta^{(3)}_2(P)$ can safely be 
neglected. It can be shown that the calculation for the regions $(2')$ 
and $(3')$ leads straightforwardly to the same conclusion. 
\par
Adding the terms of Eq.~[\ref{finalinfrarun}] and
Eq.~[{\ref{finalinfrarunpr}] where both momenta lies in the infrared, with
the parts of Eq.~[\ref{finalun}] and Eq.~[\ref{finalunpr}] with one of 
the momenta in the simple soft regime, the scale parameter $k^*$ gets 
canceled, as it should be. The divergent terms with 
'soft factors' $k>k^*$ in Eq.~[\ref{finalun}] and $q>k^*$ in 
Eq.~[\ref{finalunpr}] compensate their counterpart of
Eq.[\ref{finalsep}]. The logarithmic divergences in
Eq.~[\ref{finalinfrarun}] and Eq.~[\ref{finalinfrarunpr}] get suppressed
by the analogous term in the same Eq.~[\ref{finalsep}]. The only
remaining terms are the simple power-like divergences of
Eq.~[\ref{finalinfrarun}] and Eq.~[\ref{finalinfrarunpr}].
The result is finally the same as with the simplified spectral densities
\begin{equation}
\zeta(P) = (e^2T)^2\frac{10i}{3(2\pi)^3}\frac{1}{\mu},
\end{equation}
but the compensation of soft scale parts ($q$ or $k$ of order $eT$) had
to be proven and  the effect of the dynamical screening in the infrared
shown to be negligible. Finally, it can be mentioned that the same
cancellation of the strongest divergence occurs for three-loop diagrams,
with remaining leading order terms, although the calculations will not 
be reported here. 

\clearpage
\end{document}